\documentclass{nature} 

\usepackage{xspace}
\usepackage{color}
\usepackage{graphics}
\usepackage{upgreek}
\usepackage{amssymb}
\usepackage{amsmath}
\usepackage{siunitx}
\usepackage{booktabs}
\usepackage{mathrsfs}
\usepackage{url}
\usepackage{caption}
\usepackage[symbol]{footmisc}
\usepackage{hyperref}

\usepackage{lineno}

\usepackage{xcolor}
\definecolor{orange}{HTML}{FFA500}
\definecolor{cyan}{HTML}{48D1CC}
\definecolor{purple}{HTML}{8D32C1}
\definecolor{green}{HTML}{008000}

\def\frb{FRB~20221022A}

\def\pasa{Pubs.\ Astron.\ Soc.\ Australia}
\def\nat{Nature}
\def\aap{Astron.\ \& Astrophys.}

\def\apj{Astrophys.\ J.}
\def\apjs{Astrophys.\ J.\ Suppl.}
\def\apjl{Astrophys.\ J. Letters}

\def\mnras{Mon.\ Not.\ R.\ Astron.\ Soc.}
\def\aapr{Astron.\ \& Astrophys.\ Rev.}
\def\pasp{Publ.\ Astron.\ Soc.\ Pac.}

\newcommand{\araa}{Ann. Rev. Astron. \& Astrophys.}

\graphicspath{ {figures/} }

\title{Magnetospheric origin of a fast radio burst constrained using scintillation}

\author{Kenzie~Nimmo$^{1}$, Ziggy~Pleunis$^{2,3,4}$, Paz~Beniamini$^{5,6}$, Pawan~Kumar$^{7}$, Adam~E.~Lanman$^{1,8}$, D.~Z.~Li$^{9}$, Robert~Main$^{10,11}$, Mawson~W.~Sammons$^{10,11}$, Shion~Andrew$^{1,8}$, 
Mohit~Bhardwaj$^{12}$,
Shami~Chatterjee$^{13}$, Alice~P.~Curtin$^{10,11}$, Emmanuel~Fonseca$^{14,15}$, 
B.~M.~Gaensler$^{16,2,17}$,
Ronniy~C.~Joseph$^{10,11}$,
Zarif~Kader$^{10,11}$,
Victoria~M.~Kaspi$^{10,11}$,
Mattias~Lazda$^{16,2}$, 
Calvin~Leung$^{18,19}$,
Kiyoshi~W.~Masui$^{1,8}$, 
Ryan~Mckinven$^{10,11}$,
Daniele~Michilli$^{1,8}$,
Ayush~Pandhi$^{16,2}$, 
Aaron~B.~Pearlman$^{10,11,20,21,22}$,
Masoud~Rafiei-Ravandi$^{10,11}$,
Ketan~R.~Sand$^{10,11}$,
Kaitlyn~Shin$^{1,8}$,
Kendrick~Smith$^{23}$,
Ingrid~H.~Stairs$^{24}$
}

\begin{document}

\maketitle

\begin{affiliations}
    \item MIT Kavli Institute for Astrophysics and Space Research, Massachusetts Institute of Technology, 77 Massachusetts Ave, Cambridge, MA 02139, USA
    \item Dunlap Institute for Astronomy and Astrophysics, University of Toronto, 50 St. George Street, Toronto, ON M5S 3H4, Canada
    \item Anton Pannekoek Institute for Astronomy, University of Amsterdam, Science Park 904, 1098 XH Amsterdam, The Netherlands
    \item ASTRON, Netherlands Institute for Radio Astronomy, Oude Hoogeveensedijk 4, 7991 PD Dwingeloo, The Netherlands
    \item Department of Natural Sciences, The Open University of Israel, P.O Box 808, Ra'anana 4353701, Israel
    \item Astrophysics Research Center of the Open university (ARCO), The Open University of Israel, P.O Box 808, Ra'anana 4353701, Israel
    \item Department of Astronomy, University of Texas at Austin, Austin, TX 78712, USA
    \item Department of Physics, Massachusetts Institute of Technology, 77 Massachusetts Ave, Cambridge, MA 02139, USA
    \item Department of Astrophysical Sciences, Princeton University, Princeton, NJ 08544, USA
    \item Trottier Space Institute, McGill University, 3550 rue University, Montr\'eal, QC H3A 2A7, Canada
    \item Department of Physics, McGill University, 3600 rue University, Montr\'eal, QC H3A 2T8, Canada
    \item McWilliams Center for Cosmology, Department of Physics, Carnegie Mellon University, Pittsburgh, PA 15213, USA
    \item Cornell Center for Astrophysics and Planetary Science, Cornell University, Ithaca, NY 14853, USA
    \item Department of Physics and Astronomy, West Virginia University, PO Box 6315, Morgantown, WV 26506, USA 
    \item Center for Gravitational Waves and Cosmology, West Virginia University, Chestnut Ridge Research Building, Morgantown, WV 26505, USA
    \item Department of Astronomy and Astrophysics, University of California Santa Cruz, 1156 High Street, Santa Cruz, CA 95064, USA
    \item David A.~Dunlap Department of Astronomy \& Astrophysics, University of Toronto, 50 St.~George Street, Toronto, ON M5S 3H4, Canada
    \item Department of Astronomy, University of California Berkeley, Berkeley, CA 94720, USA
    \item NASA Hubble Fellowship Program~(NHFP) Einstein Fellow
    \item Banting Fellow
    \item McGill Space Institute Fellow
    \item FRQNT Postdoctoral Fellow
    \item Perimeter Institute for Theoretical Physics, 31 Caroline Street N, Waterloo, ON N25 2YL, Canada
    \item Department of Physics and Astronomy, University of British Columbia, 6224 Agricultural Road, Vancouver, BC V6T 1Z1 Canada
    
\end{affiliations}
\newpage
\begin{abstract}

Fast radio bursts (FRBs) are micro--to--millisecond duration radio transients\cite{2022A&ARv..30....2P} that originate mostly from extragalactic distances. The emission mechanism responsible for these high luminosity, short duration transients remains debated. The models are broadly grouped into two classes: physical processes that occur within close proximity to a central engine (e.g. Ref.\cite{2017MNRAS.468.2726K}); and central engines that release energy which moves to large radial distances and subsequently interacts with surrounding media producing radio waves (e.g. Ref.\cite{2019ApJ...872L..19M}). The expected emission region sizes are notably different between these two types of models (e.g. Ref.\,\cite{2024MNRAS.527..457K}). FRB emission size constraints can therefore be used to distinguish between these competing models and inform on the physics responsible. Here we present the measurement of two mutually coherent scintillation scales in the frequency spectrum of \frb\cite{2024arXiv240209304M}: one originating from a scattering screen located within the Milky Way, and the second originating from a scattering screen located within its host galaxy or local environment. We use the scattering media as an astrophysical lens to constrain the size of the lateral emission region\cite{2024MNRAS.527..457K}, $R_{\star\mathrm{obs}}\lesssim3\times10^{4}$\,km. We find that this is inconsistent with the expected emission sizes for the large radial distance models\cite{2020MNRAS.494.4627M}, and is more naturally explained with an emission process that operates within or just beyond the magnetosphere of a central compact object. Recently, \frb\,was found to exhibit an S-shaped polarisation angle swing\cite{2024arXiv240209304M}, supporting a magnetospheric emission process. The scintillation results presented in this work independently support this conclusion, while highlighting scintillation as a useful tool in our understanding of FRB emission physics and progenitors.

\end{abstract}

Inhomogeneities in the interstellar medium cause the radio waves from point sources to scatter, which results in temporal broadening of the signal\cite{1977ARA&A..15..479R} (parameterised by the scattering timescale $\tau_\mathrm{s}$ at some reference frequency). Scattering creates a stochastic interference pattern on the signal, called scintillation, corresponding to a frequency-dependent intensity modulation (parameterised by the characteristic frequency scale, known as the decorrelation bandwidth $\Delta \nu_\mathrm{DC}$ specified at some frequency)\cite{1977ARA&A..15..479R}. Temporal broadening becomes larger towards lower frequencies, $\tau_\mathrm{s} \propto \nu^{-\alpha}$, and spectral ``scintles'' become wider towards higher frequencies, $\Delta\nu_{\mathrm{DC}} \propto \nu^{\alpha}$. The index $\alpha$ is often close to the expectation from Gaussian density fluctuations in the scattering medium, $\alpha = 4$. Moreover, scattering and scintillation are inversely proportional\cite{1998ApJ...507..846C}: $\tau_\mathrm{s} \sim C / (2 \pi \Delta \nu_\mathrm{DC})$, with $C$ in the range 1--2.  Scattering and/or scintillation measurements in the radio signal are a powerful probe of interstellar optics\cite{1998ApJ...505..928G}. Such measurements have been used to resolve emission regions in the Crab pulsar\cite{2021ApJ...915...65M}; measure relativistic motion in Crab pulsar giant pulses\cite{2023ApJ...945..115L}; probe the circumburst environment of FRBs\cite{2015Natur.528..523M}; and have the potential to probe the structure of the circumgalactic medium (CGM; e.g. Ref.\cite{2019MNRAS.483..971V,2023arXiv230907256J}).

The CHIME/FRB experiment\cite{2018ApJ...863...48C} recently discovered the as-yet non-repeating FRB, 20221022A \cite{2024arXiv240209304M}. The event was detected by the real-time FRB search system\cite{2018ApJ...863...48C} with a S/N of 64.9 and channelised voltage data were recorded. The event was processed using the CHIME/FRB baseband pipeline\cite{2021ApJ...910..147M}, which produced a beamformed data product containing complex voltages for both X and Y polarisation hands, with a time resolution of 2.56\,$\upmu$s and frequency resolution of 0.39\,MHz. The FRB was localised\cite{2021ApJ...910..147M,2024arXiv240209304M} to equatorial coordinates Right Ascension R.A.~(J2000)\,$=\rm{03\, h\, 14\, m\, 31\, s(22)}$, Declination Dec.~(J2000)\,$=+\rm{86^{\circ}52'19''(14)}$, and associated with a host galaxy at a redshift of 0.0149(3) with posterior probability $\gtrsim99$\%. 

\begin{figure}
    \begin{center}
    \includegraphics{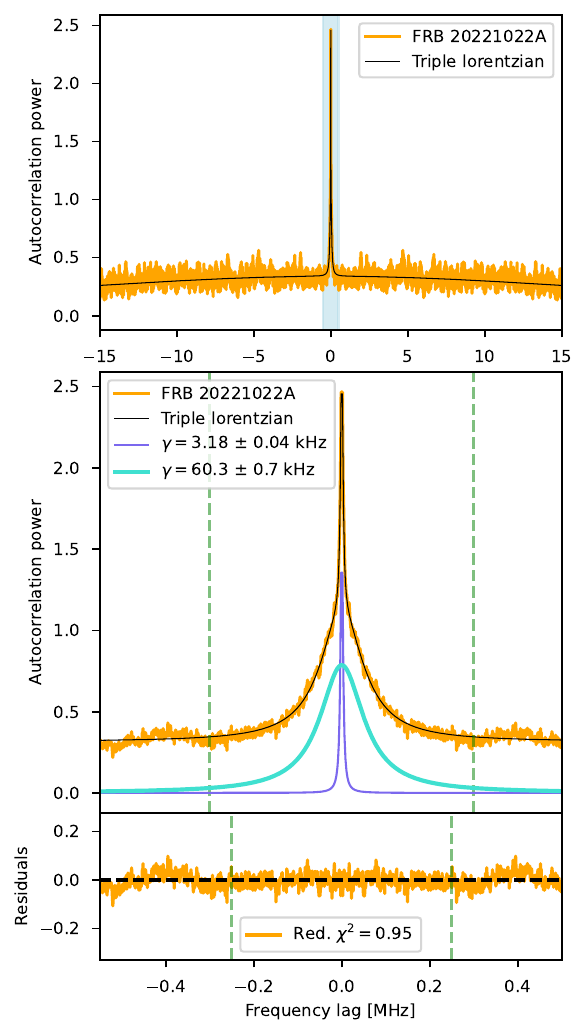}
    \end{center}
    \caption{Autocorrelation function (ACF) of the FRB~20221022A spectrum at a frequency resolution of $\sim0.76$\,kHz (orange). The top panel shows the ACF from lags $-15$\,MHz to $+15$\,MHz. The middle panel is a zoom-in on the central lag range of the ACF, highlighted by the shaded blue region on the top panel. A triple lorentzian function is fit to the ACF in the top panel between $\pm20$\,MHz and the bottom panel between $\pm0.5$\,MHz (black line; Equation~\ref{eq:lorentz}). The larger frequency scale, most clearly observable in the top panel (half-width at half-maximum $\gamma=27.3\pm0.1$\,MHz), is attributed to an instrumental ripple existing in CHIME/FRB data. The two smaller scales, which are more clearly observed in the middle zoom-in panel, are attributed to scintillation with decorrelation bandwidths of $3.18\pm0.04$\,kHz and $60.3\pm0.7$\,kHz: the individual Lorentzians are plotted on the middle panel in purple and blue, respectively. We plot the residuals in the bottom panel. The reduced-$\chi^2$ is computed within the lag-range $\pm0.25$\,MHz highlighted by the green dashed lines. We reduce the lag range since the $\sim30$\,MHz frequency scale is not expected to exhibit a Lorentzian functional form. The scintillation scales, however, are expected to be Lorentzian in form, and we find a reduced-$\chi^2$ very close to $1$ implying a good fit to the data. 
    \label{fig:entire_acf}}
    
\end{figure}

\begin{figure}
\begin{center}
    \includegraphics{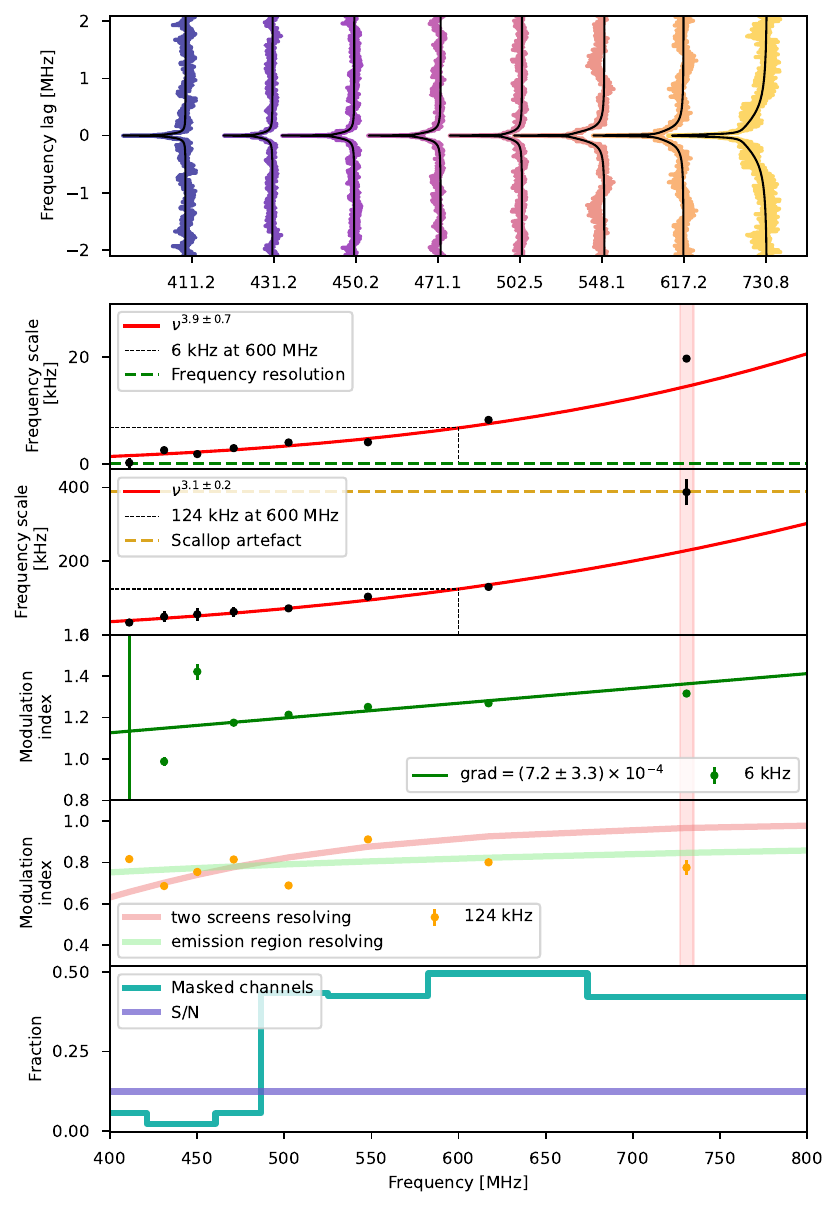}
    \end{center}
    \caption{Autocorrelation function (ACF) measured for eight subbands, containing an equal fraction of the total signal, across the CHIME observing band (top panel). The double Lorentzian fit to each ACF is shown in black. Two frequency scales are measured in each subband (shown in the second and third panels). Those frequency scales are fit with a function of the form $A\nu^{\alpha}$, shown in red. The frequency resolution is highlighted by the horizontal dashed green line, and the scallop artefact introduced into the data during the upchannelisation process is shown by the gold dashed horizontal line. The measured decorrelation bandwidths are $6\pm 1$\,kHz and $124\pm8$\,kHz at 600\,MHz, marked on the panels with the black dashed lines. We note that the high frequency data point has been omitted from all fits, indicated by the shaded red region, due to the ambiguity of the scintillation scale and the upchannelisation artefacts. The fourth and fifth panels show the modulation index measured for the $6$\,kHz and $124$\,kHz frequency scales, respectively, across the band. A least squares fit of a straight line is overplotted on the $6$\,kHz indices (forest green line), while we fit the expected evolution of the modulation index with frequency for a screen resolving the emission region size (light green line; Equation\,\ref{eq:mod_emission_resolved}; Methods) as well as the expected evolution for the screens resolving each other (pink line; Equation\,\ref{eq:mod_screen_resolved}; Methods) to the $124$\,kHz modulation indices. We note that error bars are plotted for all frequency scales and modulation indices but that they are often too small to distinguish from the marker. In the final panel, we plot the number of masked channels per subband in turquoise, and additionally the fraction of the total burst S/N per subband in purple. 
    \label{fig:sub_acf}}
\end{figure}

The beamformed baseband data were coherently and incoherently dedispersed to a DM of 116.8371\,pc\,cm$^{-3}$, measured by maximising the structure in the burst using \texttt{DM\_phase}\cite{2024arXiv240209304M,2019ascl.soft10004S}. The data were then upchannelised to a frequency resolution of $0.76$\,kHz (see Methods), at the expense of time resolution (where we are limited by the Nyquist limit: $\delta t \delta \nu \sim 1$). This frequency resolution is the highest we can achieve before diluting the signal with noise, given the total width of the FRB ($\sim 2$\,ms; Extended Data Figure\,\ref{fig:ds}). This resolution is sufficiently high to allow us to probe the expected decorrelation bandwidth from the Milky Way interstellar medium ($\sim400$\,kHz at 1\,GHz, scaling to $\sim52$\,kHz at 600\,MHz, estimated from the NE2001 Galactic electron density model\cite{2002astro.ph..7156C,2024RNAAS...8...17O}). The autocorrelation function (ACF; see Methods) of the upchannelised burst spectrum (i.e. the flux density as a function of frequency integrated over the 2-ms duration of the burst), was then computed and is shown in Figure\,\ref{fig:entire_acf}. For a burst spectrum that exhibits intensity fluctuations due to scintillation, the expected functional form of the ACF is a Lorentzian where the half-width at half-maximum is the decorrelation bandwidth\cite{1990ARA&A..28..561R}. Additionally, the ACF that we compute is normalised such that the peak of the Lorentzian is the square of the modulation index (see Methods), defined as the standard deviation of the observed spectrum divided by its mean\cite{1990ARA&A..28..561R}. Three distinct frequency scales are evident in the ACF (Figure\,\ref{fig:entire_acf}), which we measure by performing a least-squares fit of the addition of three Lorentzian functions to the ACF (see Methods). The $\sim30$\,MHz scale, which is also apparent in the burst dynamic spectrum (Extended Data Figure\,\ref{fig:ds}), is not scintillation, but rather an instrumental ripple introduced by reflections between the mesh and the focal line (separated by 5\,m) of the semi-cylindrical CHIME reflectors\cite{2022ApJS..261...29C}. We confirm that the other two frequency scales are both scintillation from two distinct scattering screens by computing the ACF for eight subbands across the CHIME observing band of $400$--$800$\,MHz, containing an equal fraction of the burst S/N, measuring both scales in each subband, and observing that they evolve with frequency with index $\alpha=3.9\pm0.7$ and $\alpha=3.1\pm0.2$ for the frequency scales $\Delta \nu_\mathrm{DC}=6\pm1$\,kHz and $\Delta \nu_\mathrm{DC}=124\pm7$\,kHz at $600$\,MHz, respectively (Figure\,\ref{fig:sub_acf}; see Methods). Scaling our scintillation measurements to 1\,GHz using the measured $\alpha$, we are able to compare with the scattering prediction from the Galactic electron density model, NE2001 ($\tau_\mathrm{s}$ at 1\,GHz)\cite{2002astro.ph..7156C}. We find that the $6$\,kHz scintillation scale is a factor of $\sim9$ less than the prediction, while the $124$\,kHz scale is a factor of $\sim1.5$ larger. Naively, one might expect the $124$\,kHz to be the Galactic scintillation scale due to its better agreement with predictions, however it is worth noting that Galactic electron density models have large uncertainties (e.g. as discussed in Ref.\cite{2008PASA...25..184G}), especially for lines-of-sight at high galactic latitude, as is the case for \frb~($b \approx 24.6^\circ$).

In addition to the decorrelation bandwidth, we also measure the modulation index of the two scintillation scales: $m_{6\,\mathrm{kHz}}=1.2\pm0.1$ and $m_{124\,\mathrm{kHz}}=0.78\pm0.07$, for the 6\,kHz and 124\,kHz scintillation scales, respectively (see Methods). Over the observing band the modulation index of both scintillation scales are consistent with being constant with frequency (Figure\,\ref{fig:sub_acf}). 

The modulation index $m_{6\,\mathrm{kHz}}$ is consistent with order unity, which indicates ``perfect" modulation from a point source. We observe that $m_{124\,\mathrm{kHz}} < m_{6\,\mathrm{kHz}}$, which one can explain by either the screen closest to the observer partially resolving the farther screen, or the farther screen partially resolving the source. In the Methods we derive the expected frequency dependence of the modulation index for both situations above, and we fit to the data in Figure\,\ref{fig:sub_acf}. For the case where the screen closest to the observer is partially resolving the farther screen, we derive a stronger frequency dependence on the modulation index (Equation\,\ref{eq:mod_screen_resolved} in Methods) than what we observe (Figure\,\ref{fig:sub_acf}), leading us to conclude that the lower modulation index, $m_{124\,\mathrm{kHz}}$, is due to the emission region being partially resolved by the farther screen (which has a weaker frequency dependence; Equation\,\ref{eq:mod_emission_resolved} in Methods). However, neither fit perfectly represents the data, which could be due to the functional forms (Equations\,\ref{eq:mod_screen_resolved} and \ref{eq:mod_emission_resolved} in Methods) becoming more complex when invoking complicated morphological structure of the scattering medium (see Methods and Figure\,\ref{fig:sub_acf}). Given that neither fit above describes the data, we strengthen the claim that the emission site is being resolved using the measured frequency dependence of the 124\,kHz scintillation scale, $\alpha=3.1\pm0.2$: for the case where everything is perfectly unresolved we expect $\alpha\sim4$, for the emission region being partially resolved by the 124\,kHz scintillation screen we expect $\alpha\sim3$, while for the 6\,kHz screen partially resolving the 124\,kHz screen we would observe a shallower relation, $\alpha\sim1$ (see Methods). These arguments motivate us to place the 
124\,kHz screen closest to the FRB source, and the 6\,kHz screen closest to the observer. Still, below we also consider the case where the order of the screens is flipped, and show that it only strengthens the constraint on the FRB emission region size.

\frb\,is confirmed to be extragalactic with a host galaxy association\cite{2024arXiv240209304M} with posterior probability $\gtrsim99$\,\%. Since the data favour the emission site being partially resolved, the second screen is likely to be extragalactic. We further support this claim by showing that coherence would not be maintained for two Galactic screens under any reasonable assumptions of screen distances (see Methods). Moreover, we place the following constraint on the product of the distance between the FRB source and the extragalactic screen, $d_{\mathrm{s}_2 \star}$, and the observer to Galactic screen distance $d_{\oplus\mathrm{s}1}$: $d_{\oplus\mathrm{s}1}d_{\mathrm{s}_2 \star} \lesssim 8.8$\,kpc$^2$ (see Methods). 

\begin{figure}
\centering
\includegraphics{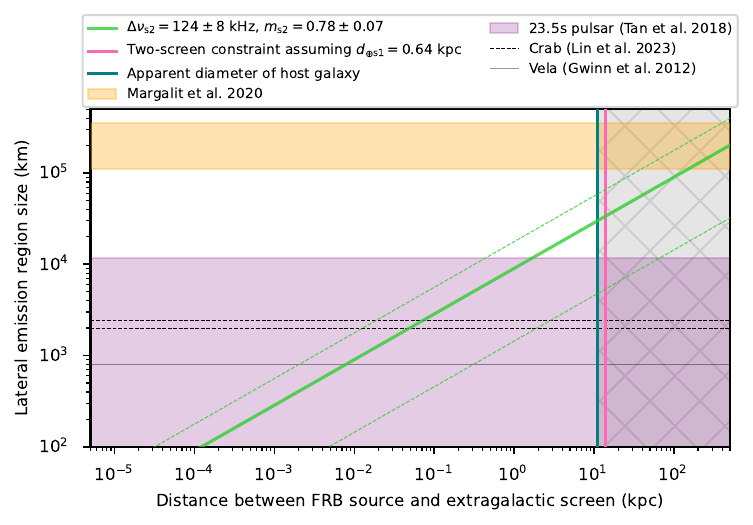}
\caption{Lateral size of the emission region as a function of the distance between the FRB and the extragalactic screen, $d_{\mathrm{s}_2\star}$. The allowable combinations of emission size and source-screen separation are shown by the green line (with the 3-$\sigma$ uncertainty region indicated by the dashed green lines). The vertical pink line indicates the two-screen constraint we apply on the source-screen distance, $d_{\mathrm{s}2\star}\lesssim14$\,kpc, assuming a Galactic screen distance of $0.64$\,kpc from NE2001\cite{2002astro.ph..7156C}. This Galactic screen distance assumption is highly uncertain, and we explore its effect on the emission size in Extended Data Figure\,\ref{fig:galscreen}. The vertical dark teal line indicates the apparent diameter of the FRB host galaxy measured in optical (11\,kpc)\cite{2003A&A...412...45P}. The grey hatched region shows the extragalactic screen distances we rule out. The orange shaded region indicates the observed emission sizes, $R_{\star\mathrm{obs}}$, inferred from the radial distances, $d$, for the synchrotron shock model determined in Ref.\cite{2020MNRAS.494.4627M} and using Equation\,\ref{eq:shock_d_R}. The purple shaded region indicates the possible emission sizes for the slowest known pulsar\cite{2018ApJ...866...54T} (which therefore has the largest magnetosphere of known radio pulsars), with horizontal black lines showing pulsar emission size measurements from the literature\cite{2012ApJ...758....7G,2023ApJ...945..115L}.  \label{fig:emission_size}
}
\end{figure}

Following the logic above, we conclude that the modulation index $m_{124\,\mathrm{kHz}} < m_{6\,\mathrm{kHz}} \sim 1$ can be explained by the emission region size being partially resolved by the extragalactic screen. This means that the angular size of the emission region projected onto the extragalactic scattering screen is slightly larger than the diffractive scale of the screen\cite{2024MNRAS.527..457K}. Naturally, this introduces a degeneracy between the emission size and screen-source distance: a larger physical emission size with a screen in the outskirts of the galaxy or a small physical emission size with a very nearby screen could result in a comparable projected angular size. Given our 124\,kHz scintillation measurement and associated modulation index we plot the allowable screen-source distance and lateral emission region size combinations in Figure\,\ref{fig:emission_size}. From the two-screen measurements, we have a constraint on the source-screen distance of $d_{\mathrm{s}_2 \star} \lesssim 14$\,kpc, which assumes a Galactic screen distance of $0.64$\,kpc from NE2001\cite{2002astro.ph..7156C} (see Methods). These two quantities are degenerate: one can increase the upper limit of $d_{\mathrm{s}_2 \star}$ by moving the Galactic screen closer to the observer. Let us consider the case where the Galactic screen is much closer to the observer, at a distance of 
$0.1$\,kpc. In this case the extragalactic screen distance can be as high as $\sim88$\,kpc. Using equations 4 and 5 from Ref.\cite{2024MNRAS.527..457K}, we estimate the electron density required to explain the scintillation measurement at this large screen distance to be $n_{e}\sim O(10^{3}\,\mathrm{cm}^{-3})$, which is at least an order of magnitude higher than the current best estimate for the Milky Way circumgalactic medium at this distance\cite{2019ApJ...880..139V}. Moreover, we would need to place \frb\,outside its host galaxy disk in order to explain why we do not measure scattering or scintillation from the disk. We therefore find that it is most plausible that the extragalactic screen is constrained within the host galaxy disk, allowing us to place the conservative constraint on $d_{\mathrm{s}_2 \star}$ from the apparent diameter of the host galaxy as observed in optical light ($\sim11$\,kpc)\cite{2003A&A...412...45P}. It is worth noting that the electron distribution extends farther than the optical diameter of the galaxy, however the inclination of the galaxy as well as the low inferred host DM\cite{2024arXiv240209304M} imply that the FRB is not traversing through the full length of the galaxy and therefore 11\,kpc is a highly conservative upper limit on the screen distance. With this upper limit on $d_{\mathrm{s}_2 \star}$, we constrain the observed emission size of $R_{\star\mathrm{obs}}\lesssim3\times10^{4}$\,km. (Figure\,\ref{fig:emission_size}). The light cylinder radius of the slowest spinning pulsar\cite{2018ApJ...866...54T}, corresponding to the largest known pulsar magnetosphere, constrains known pulsar emission region sizes to $\lesssim 10^{4}$\,km, comparable to our constraint on $R_{\star\mathrm{obs}}$ (Figure\,\ref{fig:emission_size}). 

FRB emission models are broadly characterised into two groups: magnetospheric, where the emission originates from within the magnetosphere of a compact object (e.g., Ref.\cite{2017MNRAS.468.2726K}), and non-magnetospheric, where the emission originates from much larger distances from a central compact object (e.g., Ref.\cite{2019MNRAS.485.4091M}). In the latter class of models, one can relate the lateral emission region size to the FRB emission site distance, $d$, from the central compact object\cite{2024MNRAS.527..457K}: 
\begin{equation}
\label{eq:shock_d_R}
    d \sim \frac{R_{\star\mathrm{obs}}^2}{2c\Delta t},
\end{equation}
where $\Delta t$ is the FRB temporal duration. For our upper limit on $R_{\star\mathrm{obs}}$ and $\Delta t\sim2$\,ms (Extended Data Figure\,\ref{fig:ds}), we determine an upper limit on the distance: $d<3\times10^6$\,km. For the synchrotron maser shock models, Ref.\cite{2020MNRAS.494.4627M} constrains the radial distances to be $10^7$--$10^8$\,km, which exceeds our upper limit on the distance. In order to exceed their lower bound of $10^7$\,km, we would require an FRB source to extragalactic screen distance of $>148$\,kpc, which is well beyond the apparent diameter of \frb's host galaxy. We therefore find our observations to be more consistent with the magnetospheric class of emission models\cite{2024MNRAS.527..457K} or an emission region just beyond the neutron star light cylinder radius\cite{2019ApJ...876L...6P,2023ApJ...945..115L}. Our findings independently support the conclusions drawn on \frb\, in Ref.\cite{2024arXiv240209304M}. There they observe an S-shaped polarisation position angle swing across the burst duration, often seen in pulsar pulses and attributed to an emission beam sweeping across the line of sight, indicative of a magnetospheric origin of the emission.

If we assume an emission size typical for pulsar emission, $100$--$1000$\,km,\cite{2012ApJ...758....7G,2023ApJ...945..115L} we infer an extragalactic screen distance of $\sim0.1$--$12$\,pc (Figure\,\ref{fig:emission_size}), comparable in scale to the size of the Crab nebula\cite{2008ARA&A..46..127H}. Two repeating FRBs in the literature have been observed associated with compact persistent radio sources (PRSs): FRB~20121102A\cite{2017Natur.541...58C,2017ApJ...834L...8M} and FRB~20190520B\cite{2022Natur.606..873N,2023ApJ...958L..19B}, hypothesised to be magnetised nebulae surrounding the FRB progenitor\cite{2018ApJ...868L...4M}. Motivated by the possibility of a nebula surrounding \frb\,from our scintillation measurements, we conducted follow-up observations with the European Very Long Baseline Interferometry Network (EVN; see Methods) at 1.6\,GHz. We did not detect any compact radio emission coincident with the FRB position or the FRB host galaxy down to a luminosity of $L_{1.6\mathrm{GHz}}=2\times10^{27}$\,erg\,s$^{-1}$\,Hz$^{-1}$. With our sensitivity we could have detected an FRB~20121102A-like PRS ($L_{1.7\mathrm{GHz}}=2.1\times10^{29}$\,erg\,s$^{-1}$\,Hz$^{-1}$)\cite{} with a significance of $\sim650\sigma$, and an FRB~20190520B-like PRS ($L_{3\mathrm{GHz}}=3.8\times10^{29}$\,erg\,s$^{-1}$\,Hz$^{-1}$)\cite{} with a significance of $\sim1100\sigma$. Our upper limit is in agreement with the proposed PRS luminosity - rotation measure relation\cite{2020ApJ...895....7Y}, given the relatively low measured rotation measure for \frb\,($\mathrm{RM}=-40$\,rad\,m$^{-2}$)\cite{2024arXiv240209304M}. An alternative hypothesis for the small screen distance could be that the FRB source is embedded in an HII region causing the scattering\cite{2024MNRAS.527.7568O}. We note that although pulsars typically have emission sizes $100$-$1000$\,km, smaller emission region sizes $<100$\,km as well as 1000\,km up to $3\times10^{4}$\,km are also possible for \frb\,given our scintillation measurements (Figure\,\ref{fig:emission_size}). 

In this study we independently support a magnetospheric emission model for \frb\,by constraining the lateral emission region size using the measurement of extragalactic scintillation. This highlights incredible potential for similar studies in the future to not only explore the emission physics of FRBs\cite{2024MNRAS.527..457K}, but also to explore the properties of the FRB immediate environments, which hold valuable clues into their sources and progenitors.

\clearpage

\clearpage

\clearpage



\newpage

\begin{methods}

\subsection{Scintillation analysis}
To measure scintillation, the coherently dedispersed baseband data are first upchannelised to a frequency resolution of 0.76\,kHz. The upchannelisation process is as follows: first the complex voltage dynamic spectrum is divided into time blocks of length 512 bins; for each polarisation hand, frequency channel and time block, a Fast Fourier Transform (FFT) is performed, creating an array that has a single polarisation hand, a single time bin and 512 frequency channels; the result is a complex voltage dynamic spectrum with 2 polarisation hands, $2.56\times 512$\,$\upmu$s time resolution, and $0.390625/512$\,MHz frequency resolution. This frequency resolution was chosen in order to probe the expected decorrelation bandwidth from the Milky Way interstellar medium, estimated using NE2001 ($\sim44$\,kHz at 600\,MHz; using $\tau_{\mathrm{scat},1\,\text{GHz}}\sim 0.46$\,$\upmu$s from Ref.\cite{2002astro.ph..7156C,2024RNAAS...8...17O} and the relationship $\tau_\mathrm{scat}\sim1/(2\pi\Delta\nu_{\text{DC}})$). 

In the upchannelised data product, the time resolution is sufficiently coarse such that the burst is unresolved in time. The on-burst spectrum is then taken to be the maximum S/N time bin. An off-burst spectrum is also computed for calibration purposes. The FFT used to upchannelise the baseband data introduces a scalloping artefact that repeats every 0.390625\,MHz (i.e., the width of the channels of the original channelisation of the baseband data, 400\,MHz over 1,024 channels). To correct for this artefact, the off-burst spectrum is folded to determine an average 0.390625\,MHz scallop shape, which is then divided out from the on-burst spectrum (Extended Data Figure~\ref{fig:spec}). We attribute channels in the off-burst spectrum that exceed a S/N of $3$ to radio frequency interference (RFI), and we mask both the on-burst and off-burst spectra. The autocorrelation functions (ACF) of both the on-burst and off-burst spectra are then computed using
\begin{equation}
\label{eq:acf}
{\rm ACF}(\Delta \nu) = \frac{\sum\limits_{i}(S(\nu_{i})-\bar{S})(S(\nu_{i}+\Delta\nu)-\bar{S})}{N_{\Delta\nu}(\bar{S}-\bar{S}_{\rm noise}^2)^2}, 
\end{equation}
following Ref.\cite{2019ApJ...872L..19M}. We only sum over indices $i$ that give non-masked values for $S(\nu_{i})$ and $S(\nu_{i}+\Delta\nu)$ at a given $i$. $N_{\Delta\nu}$ is the total number of unmasked overlapping frequency channels that are used to compute the ACF for a given frequency lag $\Delta\nu$. The ACF calculated using Equation~\ref{eq:acf} is normalised such that the amplitude of a characteristic frequency scale present in the ACF is the square of the modulation index of that frequency scale, where the modulation index is defined as the standard deviation of the observed burst spectrum divided by its mean\cite{1990ARA&A..28..561R}.

In Figure~\ref{fig:entire_acf} we show the on-burst ACF for FRB~20221022A for the entire observing bandwidth, with the zero lag noise spike masked and with three clear frequency scales visible by eye. There is a $27.3$\,MHz frequency scale arising from CHIME's instrumental design\cite{2018ApJ...863...48C}, which we can see by eye in the dynamic spectrum (Extended Data Figure~\ref{fig:ds}). We fit the ACF out to a lag of 20\,MHz with a triple Lorentzian function
\begin{equation}
\label{eq:lorentz}
    \frac{m_{1}^2}{1+(\Delta\nu/\gamma_{1})^2} + \frac{m_{2}^2}{1+(\Delta\nu/\gamma_{2})^2} + \frac{m_{3}^2}{1+(\Delta\nu/\gamma_{3})^2},
\end{equation}
for frequency lag $\Delta\nu$, the Lorentzian half-width at half-maximum $\gamma_{i}$ and modulation index $m_{i}$. We note that a Lorentzian is the expected functional form of the ACF, with the decorrelation bandwidth defined as the half-width at half-maximum of the Lorentzian, in order to mathematically obtain a temporal exponential decay from scatter broadening\cite{1990ARA&A..28..561R}.\footnote{Though quasi-periodic spectral structure was observed in spectra of FRB~20121102A and was suggested to arise from diffractive lensing\cite{2022ApJ...925..109L}.} Since we know that the instrumental ripple is not scintillation, we do not necessarily expect that it should adopt the functional form of a Lorentzian. The exact functional form we fit to the instrumental ripple scale is unimportant as long as we capture the amplitude of the modulation. This is because the frequency scale is orders of magnitude different from the other two scales evident in the ACF, and so only the amplitude of the modulation at the frequency lags relevant for the smaller frequency scales (i.e. around the peak) is important to return reliable modulation indices (the difference between e.g a Lorentzian or Gaussian at such small frequency lags is indistinguishable). We consider correlated uncertainties in the ACF, following Ref.\cite{1991tstm.book.....B} and the implementation in {\tt scintools}\cite{2020ApJ...904..104R}, which are propagated into the fitting procedure. We compute the reduced-$\chi^2$ statistic between lags $\pm0.25$\,MHz, relevant for the two smaller frequency scales which likely could be scintillation, and find that three Lorentzians fit well to the data with a reduced-$\chi^2$ of $0.95$. The potential decorrelation bandwidths, defined as the half-width at half-maximum of the Lorentzian, are measured to be $3.18\pm0.04$\,kHz and $60.3\pm0.7$\,kHz, with modulation indices m\,$\sim1.3$ and m\,$\sim0.89$, respectively (Figure\,\ref{fig:entire_acf}; i.e. in the ACF of the full CHIME bandwidth). We note that these frequency scales are smaller than the decorrelation bandwidths we measure in the frequency-resolved ACFs (see following paragraph and Figure\,\ref{fig:sub_acf}) due to the burst having a larger S/N in the lower half of the band (where the decorrelation bandwidth is smaller): i.e. these values are S/N-weighted.

Residual upchannelisation artefacts as well as RFI can introduce misleading frequency structure in the spectrum ACF. In order to test that the frequency scales that we measure in the on-burst ACF across the entire CHIME band are consistent with scintillation, we divide the 400\,MHz total bandwidth into eight subbands, containing an equal fraction of the burst S/N, and compute the ACF per subband to explore the frequency dependence of the putative scintillation. We measure both frequency scales in all eight subbands using a double Lorentzian fit per subband (Figure~\ref{fig:sub_acf}). The uncertainties on the ACF fit parameters are a quadrature sum of the fit uncertainties with the finite-scintle error, following the implementation in {\tt scintools}\cite{2020ApJ...904..104R}. As shown in Figure~\ref{fig:sub_acf}, we perform a least-squares fit of a function of the form $A\nu^{\alpha}$ to the half-width at half-maxima measured from the two Lorentzians fit to the ACF, for constant $A$ and index $\alpha$. To confirm the frequency scales observed are scintillation, we expect $\alpha\sim4$, while an instrumental artefact or RFI should not evolve with frequency in the same manner. Note that we omit the high frequency data point, since it cannot be distinguished from the 390\,kHz upchannelisation artefact (Figure\,\ref{fig:sub_acf}). For the smaller frequency scale, we measure $\alpha=3.9\pm0.7$, and for the larger scale we measure $\alpha=3.1\pm0.2$. The 6\,kHz frequency scale exhibits a frequency dependence consistent with the $\nu^{4}$ scaling for refractive scattering, while the 124\,kHz scale's frequency dependence is shallower (but within the range observed for pulsar scintillation\cite{2017MNRAS.470.2659G}). We therefore attribute both scales to scintillation from two scattering screens along the line of sight from FRB~20221022A to the observer. We report decorrelation bandwidths of $6\pm 1$\,kHz and $124\pm8$\,kHz at 600\,MHz, which we measure from the $A\nu^{\alpha}$ fits, and with uncertainties determined using the standard deviation of the fit residuals. The NE2001 decorrelation bandwidth prediction\cite{2002astro.ph..7156C,2024RNAAS...8...17O} at 1\,GHz in this line of sight is $\sim400$\,kHz. We scale our decorrelation bandwidths using the measured frequency scaling index $\alpha$, giving $\Delta\nu_{\text{DC, 1\,GHz}}\sim44$\,kHz, and $\Delta\nu_{\text{DC, 1\,GHz}}\sim604$\,kHz. The 6\,kHz and 124\,kHz decorrelation bandwidths are a factor of $\sim9$ lower and $\sim1.5$ higher than the NE2001 prediction, respectively.

In addition to measuring the modulation index in the entire band ACF, we also measure the modulation indices across the burst profile in time and across the observing band. In Extended Data Figure~\ref{fig:ds} we plot the modulation index measured across the burst profile in time bins of width $164\,\upmu$s. These modulation indices are measured by computing the ACF of the spectra (averaged over $164\,\upmu$s of time) with frequency resolution $24$\,kHz, and taking the square root of the peak subtracting a constant offset (introduced by the instrumental ripple). We choose this frequency resolution in order to ensure the $6$\,kHz frequency scale is unresolved and reducing its influence on the modulation index measurements. Note that in Extended Data Figure~\ref{fig:ds}, we only plot the modulation index measurements where the S/N within the 164\,$\upmu$s time interval was $>8$. The modulation index broadly appears to be constant over the burst duration, with a mean of $0.76\pm0.06$. 

\subsection{Two-screen constraints}
We consider a two-screen system as shown in Extended Data Figure~\ref{fig:twoscreen}, with the observer, $\oplus$, an astrophysical point source (here, \frb), $\star$, and two screens: $\mathrm{s}_1$ (closest to the observer) and $\mathrm{s}_2$ (closest to the source). We are following the formalism derived in Refs.\cite{2015Natur.528..523M,2022ApJ...931...87O} for an extragalactic source, but deriving it generally to allow for a Galactic source  (see e.g. Ref.\cite{2023MNRAS.525.5653S}). The temporal broadening timescale of an FRB at distance $d_{\oplus\star}$, scattered by the screen $\mathrm{s}_2$ at distance $d_{\oplus \mathrm{s}_2}$ from the observer, and distance $d_{\mathrm{s}_2 \star}$ from the FRB source, is 
\begin{equation}
    \tau_{\mathrm{s}_2} = \frac{\theta_{\mathrm{s}_2}^2}{c}\frac{d_{\oplus\star}d_{\oplus \mathrm{s}_2}}{d_{\mathrm{s}_2 \star}},
\end{equation}
where $\theta_{\mathrm{s}_2}$ is the angular-broadened size of the FRB scattered by screen $\mathrm{s}_2$ and $c$ is the speed of light\cite{2022ApJ...931...87O}. 
The coherence length of the radio waves incident on screen $\mathrm{s}_1$ is
\begin{equation}
l_{c} \simeq \frac{\lambda}{2\pi \theta_{\mathrm{s}2}} = \frac{\lambda}{2\pi}\sqrt{\frac{d_{\oplus\star}d_{\oplus \mathrm{s}_2}}{\tau_{\mathrm{s}_2}\,c\, d_{\mathrm{s}_2 \star}}}, 
\end{equation}
for observing wavelength $\lambda$. 
Scattering from screen $\mathrm{s}_2$ can weaken scintillation from $\mathrm{s}_1$ if the coherence length is reduced below the size of the Galactic scattering projected onto $\mathrm{s}_1$:
\begin{equation}
\label{eq:lcone}
    l_{\text{cone}}\simeq d_{\oplus \mathrm{s}_1}\theta_{\mathrm{s}_1} \simeq \sqrt{\tau_{\mathrm{s}_1}\,c \frac{d_{\mathrm{s}_1 \star}d_{\oplus \mathrm{s}_1}}{d_{\oplus\star}}}.
\end{equation}
With a measurement of scattering or scintillation (at least one scintillation scale is required) from both screens in the two-screen system, this sets the condition that $l_{c}\gtrsim l_{\text{cone}}$, yielding: 
\begin{equation}
    \tau_{\mathrm{s}_1}\tau_{\mathrm{s}_2} \lesssim \frac{1}{(2\pi\nu)^2} \frac{d_{\oplus\star}^2d_{\oplus \mathrm{s}_2}}{d_{\mathrm{s}_1\star}d_{\mathrm{s}_2\star}d_{\oplus \mathrm{s}_1}}.
\end{equation}

Using the relation between the scatter-broadening timescale, $\tau$, and decorrelation bandwidth from scintillation $\Delta\nu_{\mathrm{DC}}$: $\tau = \frac{C}{2\pi\Delta\nu_{\mathrm{DC}}}$ with $C\sim1$--$2$, we derive the general two-screen equation:

\begin{equation}
\label{eq:general_two}
    \Delta\nu_{\mathrm{s}_1}\Delta\nu_{\mathrm{s}_2} \gtrsim C_{\mathrm{s}_1}C_{\mathrm{s}_2}\nu^2 \frac{d_{\mathrm{s}_1\star}d_{\mathrm{s}_2\star}d_{\oplus \mathrm{s}_1}}{d_{\oplus\star}^2d_{\oplus \mathrm{s}_2}}.
\end{equation}

The high posterior probability ($>99$\,\%)\cite{2024arXiv240209304M} of the host galaxy association, confirms that \frb~ is extragalactic. We must consider if the two screens we observe are both Galactic, or if one of the screens is extragalactic. With our two measured scintillation scales in hand, we consider both of these cases below.
\begin{center}\textit{Case 1: One Extragalactic Screen and One Galactic Screen}\end{center}

First, let us assume that the screen $\mathrm{s}_2$ is extragalactic, and $\mathrm{s}_1$ is a screen within the Milky Way. In this situation we have the approximations  
\begin{equation}
    d_{\oplus\star} \simeq d_{\oplus \mathrm{s}_2} \simeq d_{\mathrm{s}_1\star} 
\end{equation}
and so we can simplify Equation~\ref{eq:general_two} to
\begin{equation} 
 \label{eq:2scr}
 \Delta\nu_{\mathrm{s}_1}\Delta\nu_{\mathrm{s}_2} \gtrsim C_{\mathrm{s}_1}C_{\mathrm{s}_2}\nu^2\frac{d_{\mathrm{s}_2\star}d_{\oplus \mathrm{s}_1}}{d_{\oplus\star}^2}.
\end{equation}
Note that typically there is a $(1+z)$ factor here\cite{2013ApJ...776..125M}, which we do not include since the redshift of \frb\,is sufficiently small ($z=0.0149$)\cite{2024arXiv240209304M} that it does not affect the results.

Given our scintillation measurements for \frb: 6\,kHz and 124\,kHz, assuming $C_{\mathrm{s}_1}=C_{\mathrm{s}_2}=1$, which is the most conservative value in this case, and taking the distance to the identified host galaxy in Ref.\cite{2024arXiv240209304M}, $d_{\oplus\star}=65.189$\,Mpc, we get the constraint:
\begin{equation}
\label{eq:screen_constraint}
d_{\oplus \mathrm{s}_1}d_{\mathrm{s}_2 \star} \lesssim 8.8\,\text{kpc}^2
\end{equation}
Using NE2001\cite{2002astro.ph..7156C,2024RNAAS...8...17O}, we can estimate $d_{\oplus \mathrm{s}_1}$ from the distance where the wavenumber spectral coefficient $C_{n}^{2}$ peaks (which can be thought of as a quantity resembling the amount of turbulence): $d_{\oplus \mathrm{s}_1}\approx 0.64$\,kpc. This gives us the constraint $d_{\mathrm{s}_2 \star} \lesssim 14$\,kpc. It is worth noting that this prediction of $d_{\oplus \mathrm{s}_1}$ is highly uncertain, and we consider its impact on $d_{\mathrm{s}_2 \star}$ and ultimately our emission region size constraints later in the Methods.

Further, the decorrelation bandwidth measurement can be used to place a limit on the individual screen distances\cite{2020MNRAS.498..651B}. Starting with Equation\,47 in Ref.\cite{2020MNRAS.498..651B} and assuming Kolmogorov turbulence\cite{2004ApJ...605..759B}, we derive
\begin{equation}
\label{eq:beniaminieq47}
    \Delta\nu_{\mathrm{s}_2} \sim \pi \nu \left(\frac{l_{\text{diff}}}{R_{F}}\right)^2, 
\end{equation}
where $l_{\text{diff}}$ is the diffraction length, or the length through the screen over which the phase changes by $1$\,radian, and $R_{F} = {cd_{\mathrm{s}_2\star}}/{\nu}$ is the Fresnel radius. Equation\,19 in Ref.\cite{2022MNRAS.510.4654B} gives the relationship between $l_{\text{diff}}$ and the phase change across the screen $\phi$
\begin{equation}
    l_{\text{diff}} \sim \phi^{-6/5}\,l_{\text{max}}\left(\frac{L}{l_{\text{max}}}\right)^{3/5},
\end{equation}
for the thickness of the screen $L$ and the maximum eddy size in the scattering medium $l_{\text{max}}$. $\phi$ is directly proportional to the DM of the screen (column depth within the thickness of the screen), $\text{DM}_{\mathrm{s}_2}$, with the relationship Equation\,17 in Ref.\cite{2022MNRAS.510.4654B}
\begin{equation}
\phi = \frac{2.6\times10^{7}\,\text{DM}_{\mathrm{s}_2}}{\nu_{\text{GHz}}}.
\end{equation}
Combining all of these relationships into Equation\,\ref{eq:beniaminieq47}, we arrive at (see also Equation\,57 in Ref.\cite{2020MNRAS.498..651B}):
\begin{equation}
\text{DM}_{\mathrm{s}_2} \sim 3\times10^{4}\text{\,pc\,cm}^{-3}\text{  } \Delta\nu_{\mathrm{s}_2}^{-5/12}\text{  } \nu_{\text{GHz}}^{11/6} \left(\frac{d_{\mathrm{s}_2\star}}{1\text{pc}}\right)^{5/12} \left(\frac{l_{\text{max}}}{L}\right)^{1/3} \left(\frac{L}{d}\right)^{5/6}.  
\end{equation}
Substituting in our measured decorrelation bandwidth $\Delta\nu_{\mathrm{s}_2} = 124$\,kHz, observing frequency $\nu_{\text{GHz}} = 0.6$, and taking the ratio of maximum eddy size over screen size to be ${l_{\text{max}}}/{L}\sim10^{-4}$ (consistent with what is seen from Milky Way turbulence):
\begin{align*}
\text{DM}_{\mathrm{s}_2} \sim 4\text{\,pc\,cm}^{-3}\text{  } \left(\frac{d_{\mathrm{s}_2\star}}{1\text{pc}}\right)^{5/12} \left(\frac{L}{d_{\mathrm{s}_2\star}}\right)^{5/6} .
\end{align*}
The contribution of the total DM attributed to the host galaxy was estimated in Ref.\cite{2024arXiv240209304M} as DM$_{\text{host}} \lesssim 14^{+23}_{-14}$\,pc\,cm$^{-3}$. We therefore estimate the following:
\begin{align*}
    (0\mathrm{-}37)\,\text{pc\,cm}^{-3} \gtrsim \text{DM}_{\mathrm{s}_2} \sim 4\text{\,pc\,cm}^{-3}\text{  } \left(\frac{d_{\mathrm{s}_2\star}}{1\text{pc}}\right)^{5/12} \left(\frac{L}{d_{\mathrm{s}_2\star}}\right)^{5/6} 
\end{align*} 
and so
\begin{align*}
\left(\frac{d_{\mathrm{s}_2\star}}{1\text{pc}}\right) \lesssim 210\,\text{pc}  \left(\frac{L}{d_{\mathrm{s}_2\star}}\right)^{-2}. 
\end{align*}

If we assume that $\frac{L}{d_{\mathrm{s}_2\star}}\sim1$, we have a tight constraint on $d_{\mathrm{s}_2\star} < 210$\,pc. However, $\frac{L}{d_{\mathrm{s}_2\star}}\sim1$ is not always a fair assumption, with values inferred $\ll 1$ for some pulsars\cite{2008MNRAS.388.1214W,2010ApJ...708..232B,2024ApJ...962...57S}. This therefore, unfortunately, does not tightly constrain the distance $d_{\mathrm{s}_2\star}$. 
\begin{center}\textit{Case 2: Two Galactic screens} \end{center}
Now we assume that the source is extragalactic, at a distance\cite{2024arXiv240209304M} of $d_{\oplus\star}=65.189$\,Mpc, but both screens $\mathrm{s}_1$ and $\mathrm{s}_2$ are within the Milky Way. Given this situation, we can make the approximations:
\begin{align*}
d_{\mathrm{s}_1\star}\simeq d_{\mathrm{s}_2\star} \simeq d_{\oplus\star}.
\end{align*}
Under this approximation, the assumption that $C_{\mathrm{s}_1}=C_{\mathrm{s}_2}=1$ and using our decorrelation bandwidth measurements, Equation\,\ref{eq:general_two} gives the constraint:
\begin{equation}
\label{eq:case2}
    \frac{d_{\oplus \mathrm{s}_1}}{d_{\oplus \mathrm{s}_2}} \lesssim \frac{\Delta\nu_{\mathrm{s}_1}\Delta\nu_{\mathrm{s}_2}}{C_{\mathrm{s}_1}C_{\mathrm{s}_2}\nu^2}\sim2\times10^{-9}.
\end{equation}
Even if we force $d_{\oplus \mathrm{s}_2}$ to be the isophotal diameter of the Milky Way, $\approx27$\,kpc\cite{1998Obs...118..201G}, this restricts $d_{\oplus \mathrm{s}_1}$ to be $\lesssim 0.0001$\,pc: it is highly unlikely that there is a screen within such close proximity to us. It is worth noting that \frb\, is $\sim64^\circ$ off the ecliptic, and therefore one of the scintillation scales coming from the solar wind can be easily ruled out. If we change $d_{\oplus \mathrm{s}_2}$ to be smaller, the condition in Equation\,\ref{eq:case2} forces $d_{\oplus \mathrm{s}_1}$ to be even smaller, supporting that this outcome is highly unlikely. 

We note that if we consider the case where both screens are extragalactic, the problem is symmetric and the same constraint applies. Suppose the farther screen is $50$\,kpc from the source, out in the host galaxies halo, then the nearby screen would need to be $<0.0001$\,pc. While pulsars are known to scintillate from bow shocks very close to the source\cite{2024MNRAS.527.7568O}, this configuration is much more fine-tuned and therefore more unlikely than the case where one of the screens is Galactic. 

Throughout this section we have implicitly assumed that the screens are two-dimensional and isotropic. The ACF in Figure\,\ref{fig:entire_acf} is well-fit with a double Lorentzian function. We therefore find no deviations from the expectations of the isotropic screen assumption. Deviations from these expectations, however, can be subtle, and so we explore below the possibility of one-dimensional anisotropic screens and the implications for our conclusions. 

\subsection{One-dimensional, anisotropic screens}

Throughout this manuscript, the implicit assumption we make is that the scintillation screens are isotropic and two-dimensional. This assumption means that the angular broadening of the source due to the screen closest to the observer is equivalent to the size of the source as seen by the farther screen. However, if the screens are sheet-like\cite{2023arXiv230907256J} (i.e. the normal vector of the ``sheet" is perpendicular to the line of sight, rather than parallel in the case of the thin-screen model), the angular broadening is direction-dependent, introducing a dependence on the angle between the one-dimensional screens. The condition $l_{c}\gtrsim l_{\text{cone}}$ from the sub-section above, becomes $l_{c}\gtrsim l_{\text{cone}}\mathrm{cos}(\theta)$, where $\theta$ is the angle between the two sheet-like screens projected onto the line-of-sight plane. \\
For Case 2 described above, Equation\,\ref{eq:case2} becomes
\begin{equation}
    \frac{d_{\oplus \mathrm{s}_1}}{d_{\oplus \mathrm{s}_2}} \mathrm{cos}^2(\theta) \lesssim \frac{\Delta\nu_{\mathrm{s}_1}\Delta\nu_{\mathrm{s}_2}}{C_{\mathrm{s}_1}C_{\mathrm{s}_2}\nu^2}\sim2\times10^{-9}.
\end{equation}
For reasonable $d_{\oplus \mathrm{s}_1}$ and $d_{\oplus \mathrm{s}_2}$, this inequality can be satisfied by invoking a $\mathrm{cos}(\theta) \ll 1$, or equivalently making the one-dimensional screens almost perfectly perpendicular. This is very tightly constraining the geometry of the scattering media which is fine-tuned in reality and therefore unrealistic. Additionally, as discussed in the following section, for the larger scintillation scale, with modulation index $<1$, we find the decorrelation bandwidth and modulation index frequency dependence to agree more with the emission size being resolved than the screens resolving each other. These frequency dependencies are not affected by the $\mathrm{cos}(\theta)$ term and therefore add further doubt to the scenario of an extragalactic source with two almost perpendicular one-dimensional Galactic screens. \\
In Ref.\cite{2023arXiv230907256J}, it is shown that one can observe a suppression of the modulation index for the larger scintillation scale if the finer scintillation scale is unresolved by the telescope frequency resolution. However, this situation does not apply to this work since we have resolved both scintillation scales in our analysis.\\

\subsection{Suppressed intensity modulation}

The case studies presented above support the extragalactic nature of the second screen, $\mathrm{s}_2$. The two-screen constraints in Equation\,\ref{eq:screen_constraint} place the second screen likely within the host galaxy. We observe no clear frequency or time evolution of the modulation index of the 124\,kHz scintillation scale (Figure\,\ref{fig:sub_acf}, Extended Data Figure\,\ref{fig:ds}). The modulation index for \frb\,was observed to decrease over the burst profile (which is dominated by an exponential scattering tail) due to the two screens partially resolving each other\cite{2023MNRAS.525.5653S}. In the case presented here, we are not resolving the scattering timescale, and so it is not surprising that we do not observe an evolution of the modulation index with time. We explore the possibility that the modulation index $m_{124\,\mathrm{kHz}}<1$ observed is either due to the screens resolving each other or due to the emission region size being resolved. We note that in the case of weak scintillation\cite{1990ARA&A..28..561R}, one can expect $m_{\mathrm{weak}}\sim0.1-0.3$, which is lower than our measurement of $m_{124\,\mathrm{kHz}}\sim0.78$. When the source or screen is resolved, different scintillation patterns are effectively being averaged. This has the effect of smearing the scintillation pattern in frequency and suppressing the amplitude of the intensity modulation. For this reason, in both of these cases we expect different modulation index and decorrelation bandwidth frequency dependencies, which we derive below. \\
First we derive the relationship for the case where the observed emission region size is being partially resolved.\\
The physical size of the extragalactic screen, $\mathrm{s}_2$, is 
\begin{equation}
\label{eq:s2size}
L_{\mathrm{s}_2} = \theta_{\mathrm{s}_2}d_{\mathrm{s}_2\star},
\end{equation}
where $\theta_{\mathrm{s}_2}$ is the angular size of screen $\mathrm{s}_2$ from the perspective of the FRB source, and 
\begin{equation}
\label{eq:s2angle}
\theta_{\mathrm{s}_2} = \sqrt{\frac{2c\tau_{\mathrm{s}_2}}{d_{\mathrm{s}_2\star}}} = \sqrt{\frac{c}{\pi\Delta\nu_{\mathrm{s}_2}d_{\mathrm{s}_2\star}}},
\end{equation}
where we relate the scattering timescale and decorrelation bandwidth through the relation $\tau_{\mathrm{s}_2}\sim1/(2\pi\Delta\nu_{\mathrm{s}_2})$. Substituting Equation\,\ref{eq:s2angle} into Equation\,\ref{eq:s2size} yields:
\begin{equation}
\label{eq:l_s2}
L_{\mathrm{s}_2} = \sqrt{\frac{cd_{\mathrm{s}_2\star}}{\pi\Delta\nu_{\mathrm{s}_2}}}.
\end{equation}
The physical resolution of the screen is then 
\begin{equation}
    \label{eq:chi_s2}
    \chi_{\mathrm{s}_2} = \frac{1}{\sqrt{2}\pi}\frac{\lambda}{L_{\mathrm{s}_2}} d_{\mathrm{s}_2\star} = \frac{1}{\nu}\sqrt{\frac{cd_{\mathrm{s}_2\star}\Delta\nu_{\mathrm{s}_2}}{2\pi}},
\end{equation}
where the $\frac{1}{\sqrt{2}\pi}$ is a model-dependent factor\cite{1998ApJ...507..846C}. 

Substituting Equation\,\ref{eq:chi_s2} into\cite{1998ApJ...505..928G}
\begin{equation}
\label{eq:mod_gwinn}
m_{\mathrm{s}_2} = \frac{1}{\sqrt{1+4\left(\frac{R_{\star\text{obs}}}{\chi_{\mathrm{s}_2}}\right)^2}},
\end{equation}
where $R_{\star\text{obs}}$ is the observed emission region size, we derive the relationship between the lateral emission region size and the distance between the source and extragalactic screen:
\begin{equation}
\label{eq:mod_emission_resolved}
R_{\star\text{obs}} = \sqrt{\frac{c\,d_{\mathrm{s}_2\star}\,\Delta\nu_{\mathrm{s}_2}(\nu)}{8\pi\nu^2}\left(\frac{1}{m_{\mathrm{s}_2}^2} - 1 \right)}.
\end{equation}
Following a similar line of reasoning, we derive an equivalent relationship for the case where the two screens are partially resolving each other:
\begin{equation}
\label{eq:mod_screen_resolved}
    m_{\mathrm{s}2} = \frac{1}{\sqrt{1+\left(\frac{\nu}{d_{\mathrm{s}_1\mathrm{s}_2}}\right)^2\frac{8\,d_{\mathrm{s}_2\star}\,d_{\oplus \mathrm{s}_1}}{\Delta\nu_{\mathrm{s}_1}(\nu)\Delta\nu_{\mathrm{s}_2}(\nu)}}}
\end{equation}

In Figure\,\ref{fig:sub_acf} we plot the least squares fit of the modulation indices as a function of frequency with their expected relationships: Equation\,\ref{eq:mod_emission_resolved} for the partially resolved emission region size, and Equation\,\ref{eq:mod_screen_resolved} for the two-screens partially resolving each other. It is evident that in the case of the two screens resolving each other we expect a stronger frequency dependence than what is observed suggesting that the data are more in agreement with the case of the emission region being partially resolved (although neither fit describes the data with our measured reduced $\chi^2>1$: quantitatively we measure reduced $\chi_{\nu}^2\sim117$ for the resolving screens, and reduced $\chi_{\nu}^2\sim67$ for the emission region being resolved). We note that these functional forms can become more complex by invoking a complicated morphological structure of the scattering material, which is one reason why the fits may be poor. Another reason could be that the modulation index of the 124\,kHz scintillation scale is suppressed by an aspect of the analysis performed, e.g. during the upchannelisation artefact removal process. We additionally consider the case where the modulation index is $\sim1$, however, as we show later, this is less conservative for the emission region size constraints than using the $m_{\mathrm{s}2}\sim0.78$ measurement.

For both scenarios, we now derive the decorrelation bandwidth frequency dependencies. From Equation 46 in Ref.\cite{1998ApJ...505..928G}
\begin{equation}
\label{eq:t_v_gwinn}
\nu_{\mathrm{scint}} = \frac{\sqrt{1+4\sigma_1^2}}{2 \pi \tau_{\mathrm{scatt}}},
\end{equation}
where $\sigma_1 = R_{\star\rm{obs}} / \chi_{\mathrm{s}_2}$ for the case where the emission region is being resolved (see Equation\,\ref{eq:mod_gwinn}), and $\sigma_1 = L_{\mathrm{s}_2} / \chi_{\mathrm{s}_1}$ for the case where the screen is being resolved. First let us consider a partially resolved emission region. In this case, $
\chi_{\mathrm{s}_2} \propto \nu$ (see Equation\,\ref{eq:chi_s2}), which in turn means that $\sigma_1 \propto \nu^{-1}$. From Equation\,\ref{eq:t_v_gwinn}, this then gives the following frequency dependence: 
\begin{equation}
\label{eq:scint_bw_emission}
    \nu_{\rm{scint}} \propto \sqrt{A\nu^{8} + B\nu^{6}}. 
\end{equation}
In the case where the screen is being resolved, $L_{\mathrm{s}_2}\propto \nu^{-2}$ (see Equation\,\ref{eq:l_s2}), $\chi_{\mathrm{s}_1}\propto \nu$ (from Equation\,\ref{eq:chi_s2}), which then results in $\sigma_1\propto \nu^{-3}$. From Equation\,\ref{eq:t_v_gwinn}, this then gives the following frequency dependence: 
\begin{equation}
\label{eq:scint_bw_screens}
    \nu_{\rm{scint}} \propto \sqrt{C\nu^{8} + D\nu^{2}}. 
\end{equation}
For completely unresolved emission the first term in both Equations\,\ref{eq:scint_bw_emission} and \ref{eq:scint_bw_screens} dominates, and we arrive at the $\nu^{4}$ frequency scaling for the decorrelation bandwidth. However, if the scintillation is (partially) resolved, the second term dominates. For the emission region being resolved, the frequency dependence becomes $\nu_{\rm{scint}} \propto \nu^{3}$ and for the screens resolving each other we arrive at $\nu_{\rm{scint}} \propto \nu$. Our measured frequency scaling of $\alpha=3.1\pm0.2$ for the 124\,kHz scintillation scale (Figure\,\ref{fig:sub_acf}) supports that the emission region size is being partially resolved. $\alpha=4$, i.e. the case where the emission region is unresolved, is $>3\sigma$ inconsistent.

\subsection{Emission size constraints}

As outlined in Ref.\cite{2024MNRAS.527..457K}, a measurement of scintillation from a screen in the FRB host galaxy can be used to constrain the size of the FRB emission region, which in turn could be used to distinguish between FRB emission models. The $124$\,kHz modulation index frequency evolution and decorrelation bandwidth frequency relation supporting the emission region size being partially resolved, suggests that the $124$\,kHz scintillation scale is a result of the extragalactic screen, $\mathrm{s}_2$. The high reduced-$\chi^2$ of the modulation index vs frequency fit, as well as the inconsistency with the NE2001 prediction, as mentioned earlier, means we cannot rule out the scenario where neither the emission region nor screen is being partially resolved. We, therefore, consider all cases here: (a) $124$\,kHz scintillation scale from the extragalactic screen, that is partially resolving the emission region, $m_{124\,\mathrm{kHz}}=0.78$; (b) $124$\,kHz scintillation scale from the extragalactic screen, with an unresolved emission region, $m_{124\,\mathrm{kHz}}\sim1$; and (c) $6$\,kHz scintillation scale from the extragalactic screen, with an unresolved emission region, $m_{6\,\mathrm{kHz}}\sim1$.\\
In Figure\,\ref{fig:emission_size} we plot the lateral emission size as a function of the extragalactic screen distance for case (a): which is the case our data agrees with most, while also being the most conservative constraint on the emission region size. There is a clear degeneracy between the lateral emission region size and the FRB to extragalactic screen distance, which naturally arises since the $m_{\mathrm{s}2}=0.78$ measurement fixes the projected size of the emission region on the screen. As shown earlier, we have a constraint on the screen distance, $d_{\mathrm{s}_2\star}<14$\,kpc (Equation\,\ref{eq:2scr}; assuming $d_{\oplus \mathrm{s}_1}=0.64$\,kpc, from NE2001\cite{2002astro.ph..7156C}). With this limit, we can see from Figure\,\ref{fig:emission_size} that the lateral emission size upper limit is lower than the estimated size for the synchrotron maser shock model\cite{2019MNRAS.485.4091M,2020MNRAS.494.4627M}. However, this hinges on the Galactic screen distance we have assumed from the NE2001 estimate, which can be highly uncertain. We can find consistency with the shock model emission region sizes, by invoking a $d_{\oplus \mathrm{s}_1}$ that is $\sim60$\,pc, which is small but could be possible (Extended Data Figure\,\ref{fig:galscreen}). Importantly, though, in order to have an observed emission region size consistent with the shock model, the extragalactic screen distance has to be $\gtrsim148$\,kpc (Figure\,\ref{fig:emission_size}), which is significantly higher than the apparent diameter of the host galaxy ($\sim11$\,kpc)\cite{2003A&A...412...45P}. It is worth noting that this apparent diameter is derived from optical observations, while the electron distribution will extend farther, however the inclination of the galaxy with respect to the line-of-sight, as well as the low inferred host DM\cite{2024arXiv240209304M} make it highly unrealistic that \frb\,propagated through the full extent of the galactic disk, making this upper limit very conservative. It is highly unlikely that the extragalactic screen is far out in the halo of the host galaxy (or farther), but is rather from the interstellar medium of the host galaxy or the local environment of the FRB. We, therefore, place the conservative constraint on the lateral emission region size of $R_{\star\mathrm{obs}} \lesssim 3\times10^{4}$\,km.  \\
It is worth noting that there are two foreground stars\cite{2020yCat.1350....0G} at distances of $\sim0.5$\,kpc and $\sim0.8$\,kpc (broadly consistent with the $d_{\oplus \mathrm{s}_1}=0.64$\,kpc estimate from NE2001) coincident with the FRB position and host galaxy, identified in Ref.\cite{2024arXiv240209304M}. These stars could create a scintillation screen from their stellar winds, as has been observed for hot stars\cite{2017ApJ...843...15W} extending out to $\sim2$\,pc: the projected area on the sky would encompass the entire host galaxy and FRB localisation region. The two foreground stars in the \frb\,field, however are lower temperature than those observed in Ref.\cite{2017ApJ...843...15W} and so would have a lower mass loss rate and the surroundings would have a lower density. A stellar wind screen could explain the inferred larger density than the NE2001 prediction for the case where the $6$\,kHz scintillation scale is the Galactic scale, which is $\sim9$ times lower -- i.e. an $\sim9$ times higher scattering timescale -- compared with NE2001. However, without very long baseline interferometry (VLBI) to constrain the Galactic screen distance and geometry, we cannot confirm that the stellar wind is causing the Galactic scintillation here.\\
Finally, let us consider cases (b) and (c) above. In both of these cases, we assume $m_{\mathrm{s}2}\sim1$ which tells us that the emission region is a point source as viewed from the extragalactic screen. This, therefore, constrains only a minimum distance between the FRB and extragalactic screen for a given source size (Extended Data Figure\,\ref{fig:othercases}). The allowable lateral emission region size and screen distance combinations are shown on Extended Data Figure\,\ref{fig:othercases} in green and blue for case (b) and (c), respectively. In order to have an emission region size consistent with the shock model\cite{2020MNRAS.494.4627M}, we require $d_{\mathrm{s}2\star}>12$\,Mpc and $d_{\mathrm{s}2\star}>250$\,Mpc for case (b) and (c), respectively. Since the FRB is at a distance of $\sim65$\,Mpc, the non-magnetospheric model cannot work for case (c). There is no obvious nearby galaxy with a halo that could conceivably intersect the FRB line of sight, and so a scattering screen $>12$\,Mpc from the FRB is highly unlikely. Moreover, this requires a Galactic screen distance $\lesssim1$\,pc given our two-screen constraints (Equation\,\ref{eq:2scr}), which is unreasonably close, especially since \frb\,is $\sim64^\circ$ off the ecliptic, and therefore we can rule out the Galactic scintillation scale arising from the solar wind.\\
Given our observed emission region size constraints, our observations disfavour the synchrotron maser shock model\cite{2019MNRAS.485.4091M,2020MNRAS.494.4627M}. Our results are more consistent with the magnetospheric class of FRB emission models\cite{2024MNRAS.527..457K} or emission originating just beyond the light cylinder of a neutron star (e.g. Ref.\cite{2019ApJ...876L...6P,2023ApJ...945..115L}). This supports the findings of Ref.\cite{2024arXiv240209304M}, where wemeasure a polarisation angle S-shaped swing in \frb, which has been attributed to a beam sweeping across the observers line of sight, therefore tying the emission site to the rotation of an object.\\
Assuming an emission region size comparable to those observed in pulsars (100\,km--1000\,km; Ref.\cite{2012ApJ...758....7G,2023ApJ...945..115L}), motivated by the pulsar-like polarisation angle swing\cite{2024arXiv240209304M}, we infer an extragalactic screen distance from the source of $\sim0.1$--$12$\,pc (Figure\,\ref{fig:emission_size}), consistent with the size of the Crab nebula\cite{2008ARA&A..46..127H}.

\subsection{European VLBI Network imaging}
Motivated by the possibility that the scintillation scale is coming from a surrounding nebula, we observed the field of \frb\,with the European Very Long Baseline Interferometry (VLBI) Network (EVN) to search for any compact radio emission (project ID: rn002). These observations were conducted during an e-VLBI session, where the data were correlated in real-time using {\tt SFXC}\cite{2015ExA....39..259K} at the Joint Insitute for VLBI ERIC (JIVE). We observed with the EVN from 9 April 2024 22:01:55 UT to 10 April 2024 04:22:30 UT, with the following participating stations: Jodrell Bank Mark2, Effelsberg, Medicina, Noto, Onsala (On-85), Tianma (T6), Toru\'n and Irbene. The central observing frequency of our observations is 1.6\,GHz, with a bandwidth of 128\,MHz. The interferometric data were correlated with time and frequency integration of 2\,s and 0.5\,MHz, respectively. We correlated the target data at the position R.A.~(J2000)\,$=03\mathrm{h}14\mathrm{m}17.4\mathrm{s}$, Dec.~(J2000)\,$=86^\circ52'01''$, which is consistent with the centre of \frb's associated host galaxy\cite{2024arXiv240209304M}. In addition to the target scans, we observed J0217$+$7349 as the flux and bandpass calibrator, J0213$+$8717 as the phase calibrator (at a spatial separation of 0.89$^\circ$ from the pointing centre), and J0052$+$8627 as the check source. Traditional phase referencing observations were conducted with a cycle time of $\sim6.5$\,min: 5\,min on target, 1.5\,min on the phase calibrator. In total, we observed the field of \frb\,for $\sim4$\,hours. We note that we did not get target data with On-85 due to the high elevation of the source. \\
Raw voltage data are recorded from each participating telescope with  circular polarisation feeds and 2-bit sampling in VDIF\cite{whitney_2010_ivs} format. The correlated visibilities were calibrated and imaged using standard procedures in the Astronomical Image Processing System ({\tt AIPS}\cite{2003ASSL..285..109G}) and {\tt DIFMAP}\cite{1994BAAS...26..987S}. First, using the results of the automatic EVN pipeline\footnote{\href{https://evlbi.org/handling-evn-data}{https://evlbi.org/handling-evn-data}}, we performed amplitude calibration using the gain curves and individual station system temperature measurements, applied the bandpass calibration, as well as some basic flagging. We then performed some additional manual flagging of the fringe finder, before removing the instrumental delay. The final step of the calibration was to correct the phases for the entire observation, as a function of time and frequency, by performing a fringe fit using the calibrator sources. Throughout, we use Effelsberg, the most sensitive telescope in our array, as the reference antenna. \\
After calibration we image the check source to confirm that we detect it as a point source, as expected, and at the correct sky position. We then perform a grid search $\pm102$\,arcseconds around the target phase centre. This grid search comprised of making dirty maps of $\sim2\times2$\,arcseconds spanning the entire $102\times102$\,arcsecond grid, and reporting the peak of each dirty map. We make dirty maps using both natural and uniform weighting, resulting in beam sizes of $3.6\times6.9$\,mas, and $2.2\times4.6$\,mas, respectively. The resulting rms noise levels are $42\,\upmu$Jy/beam and  $63\,\upmu$Jy/beam for the natural and uniform weighted images, respectively. Given our shortest baseline (Irbene-to-Toru\'n; $\sim452$\,km), we are resolving out radio emission with size larger than approximately 82\,mas. \\
Due to time and frequency smearing we can expect to lose sensitivity as we move farther from the phase centre. Across the extent of the host galaxy we expect to lose at most 10\% of the sensitivity, while at the edge of the $1$-$\sigma$ FRB baseband localisation\cite{2024arXiv240209304M}, we lose $\sim30$\%. We did not detect any persistent compact radio emission in our search, down to a luminosity limit of $L_{1.6\,\mathrm{GHz}}<2\times10^{27}$\,erg\,s$^{-1}$\,Hz$^{-1}$ (7$\sigma$. There is a possible 6.6$\sigma$ candidate at the edge of the FRB 3$\sigma$ localisation region, that is not detected in the The Very Large Array Sky Survey (VLASS)\cite{2020PASP..132c5001L}. Confirming the astrophysical nature of this candidate is deferred to future work, but given its $\sim3\sigma$ offset from the FRB position, and large offset from the host galaxy, it seems unlikely to be related to \frb. We confirm that the NVSS source reported in Ref.\cite{2024arXiv240209304M}, NVSS J031417$+$865200, co-located with the centre of the FRB host galaxy is resolved out on our long baselines. This supports their conclusion that it is from star-formation in the host galaxy.
\subsection{Effelsberg single dish FRB search}
Although \frb\,is an as-yet non-repeating FRB, we recorded high time resolution search data with Effelsberg in parallel to search for possible repeat bursts. This search data was recorded at Effelsberg during the target scans in psrfits format using the Effelsberg Direct Digitization (EDD) backend, with time and frequency resolution 49.2\,$\upmu$s and 0.12\,MHz, respectively. The bandwidth of these data is from $1.5$ to $1.75$\,GHz, i.e. an observing band of 250\,MHz. The total intensity {\tt psrfits} data from the Effelsberg EDD backend were converted to filterbank format using {\tt digifil}\cite{2011PASA...28....1V}, conserving the time and frequency resolution of the {\tt psrfits} data. This was done in order to be compatible with {\tt Heimdall}\footnote{\href{https://sourceforge.net/projects/heimdall-astro/}{https://sourceforge.net/projects/heimdall-astro/}}, which we use for the single pulse search. Before performing the burst search we masked frequency channels that were found to contain RFI. Single pulse candidates above a S/N threshold of $7$ identified by Heimdall were then classified using FETCH (models A and H, with a probability threshold of 0.5)\cite{2020MNRAS.497.1661A}. The FETCH candidates as well as the Heimdall candidates with DMs in the range $115$--$118$\,pc\,cm$^{-3}$ were inspected by eye. We found no promising FRB candidates above a S/N of $7$. Using the radiometer equation\cite{2003ApJ...596.1142C}, taking the typical Effelsberg system temperature and gain values as $20$\,K and $1.54$\,K/Jy, respectively, and assuming a burst width of $1$\,ms, we arise at the flux upper limit of $0.1$\,Jy\,ms for this observation. Due to the sporadic activity behaviour of repeating FRBs (e.g. Ref.\cite{2022ApJ...927...59L}), our non-detection cannot confirm that \frb\,will never repeat in the future.

\subsection{Rise and Decay Times}
As discussed in Ref.\cite{2024arXiv240209304M}, the burst shows no clear evidence for temporal broadening due to multi-path propagation, with an upper limit of $\tau_{\text{scatt}}<550\,\upmu$s at 400\,MHz. The decorrelation bandwidth measurements presented in this work are consistent with this upper limit: the smallest decorrelation bandwidth, $6$\,kHz, corresponds to the larger temporal broadening scale through the relation $\tau_{\text{scatt}} \sim C / (2\pi\Delta\nu_{\text{DC}})$, which gives a scatter broadening timescale of approximately $112\,\upmu$s at $400$\,MHz. This confirms that the burst morphology is dominated by the intrinsic burst decay time, as opposed to the exponential decay from scatter broadening, as indicated by the scattering upper limits presented in Ref.\cite{2024arXiv240209304M}. Both the rise and decay times can be important quantities for probing the burst emission physics\cite{2020MNRAS.498..651B}. For example, it is difficult to explain extremely short temporal variations in the synchrotron maser model\cite{2020MNRAS.498..651B,2021NatAs...5..594N}.

We measure the rise time as the beginning of the first burst component to the peak of the first component. The peak of all three burst components are determined using the {\tt fitburst} fit described in Ref.\cite{2024arXiv240209304M}. The beginning of the first component is defined as the time from the peak that contains $90$\,\% of the fluence of the leading half of the first burst component. Similarly the decay time is computed as the time between the peak of the third component to the end of the third burst component. The end is similarly defined as the time from the peak that contains $90$\,\% of the fluence of the trailing half of the third burst component. The rise and decay times are shown in Extended Data Figure\,\ref{fig:ds}, and are measured to be $553\pm55\,\upmu$s and $492\pm49\,\upmu$s, respectively. This gives a ratio of rise/decay time of $1.1$. Measuring a ratio of rise/decay time $\ll1$ would disfavour the synchrotron shock model, however, our constraint of order unity is not constraining for emission models.


\begin{thebibliography}{10}
\expandafter\ifx\csname url\endcsname\relax
  \def\url#1{\texttt{#1}}\fi
\expandafter\ifx\csname urlprefix\endcsname\relax\def\urlprefix{URL }\fi
\providecommand{\bibinfo}[2]{#2}
\providecommand{\eprint}[2][]{\url{#2}}

\bibitem{2022A&ARv..30....2P}
\bibinfo{author}{{Petroff}, E.}, \bibinfo{author}{{Hessels}, J.~W.~T.} \& \bibinfo{author}{{Lorimer}, D.~R.}
\newblock \bibinfo{title}{{Fast radio bursts at the dawn of the 2020s}}.
\newblock \emph{\bibinfo{journal}{\aapr}} \textbf{\bibinfo{volume}{30}}, \bibinfo{pages}{2} (\bibinfo{year}{2022}).
\newblock \eprint{2107.10113}.

\bibitem{2017MNRAS.468.2726K}
\bibinfo{author}{{Kumar}, P.}, \bibinfo{author}{{Lu}, W.} \& \bibinfo{author}{{Bhattacharya}, M.}
\newblock \bibinfo{title}{{Fast radio burst source properties and curvature radiation model}}.
\newblock \emph{\bibinfo{journal}{\mnras}} \textbf{\bibinfo{volume}{468}}, \bibinfo{pages}{2726--2739} (\bibinfo{year}{2017}).
\newblock \eprint{1703.06139}.

\bibitem{2019ApJ...872L..19M}
\bibinfo{author}{{Macquart}, J.~P.} \emph{et~al.}
\newblock \bibinfo{title}{{The Spectral Properties of the Bright Fast Radio Burst Population}}.
\newblock \emph{\bibinfo{journal}{\apjl}} \textbf{\bibinfo{volume}{872}}, \bibinfo{pages}{L19} (\bibinfo{year}{2019}).
\newblock \eprint{1810.04353}.

\bibitem{2024MNRAS.527..457K}
\bibinfo{author}{{Kumar}, P.}, \bibinfo{author}{{Beniamini}, P.}, \bibinfo{author}{{Gupta}, O.} \& \bibinfo{author}{{Cordes}, J.~M.}
\newblock \bibinfo{title}{{Constraining the FRB mechanism from scintillation in the host galaxy}}.
\newblock \emph{\bibinfo{journal}{\mnras}} \textbf{\bibinfo{volume}{527}}, \bibinfo{pages}{457--470} (\bibinfo{year}{2024}).
\newblock \eprint{2307.15294}.

\bibitem{2024arXiv240209304M}
\bibinfo{author}{{Mckinven}, R.} \emph{et~al.}
\newblock \bibinfo{title}{{A pulsar-like swing in the polarisation position angle of a nearby fast radio burst}}.
\newblock \emph{\bibinfo{journal}{arXiv e-prints}} \bibinfo{pages}{arXiv:2402.09304} (\bibinfo{year}{2024}).
\newblock \eprint{2402.09304}.

\bibitem{2020MNRAS.494.4627M}
\bibinfo{author}{{Margalit}, B.}, \bibinfo{author}{{Metzger}, B.~D.} \& \bibinfo{author}{{Sironi}, L.}
\newblock \bibinfo{title}{{Constraints on the engines of fast radio bursts}}.
\newblock \emph{\bibinfo{journal}{\mnras}} \textbf{\bibinfo{volume}{494}}, \bibinfo{pages}{4627--4644} (\bibinfo{year}{2020}).
\newblock \eprint{1911.05765}.

\bibitem{1977ARA&A..15..479R}
\bibinfo{author}{{Rickett}, B.~J.}
\newblock \bibinfo{title}{{Interstellar scattering and scintillation of radio waves.}}
\newblock \emph{\bibinfo{journal}{\araa}} \textbf{\bibinfo{volume}{15}}, \bibinfo{pages}{479--504} (\bibinfo{year}{1977}).

\bibitem{1998ApJ...507..846C}
\bibinfo{author}{{Cordes}, J.~M.} \& \bibinfo{author}{{Rickett}, B.~J.}
\newblock \bibinfo{title}{{Diffractive Interstellar Scintillation Timescales and Velocities}}.
\newblock \emph{\bibinfo{journal}{\apj}} \textbf{\bibinfo{volume}{507}}, \bibinfo{pages}{846--860} (\bibinfo{year}{1998}).

\bibitem{1998ApJ...505..928G}
\bibinfo{author}{{Gwinn}, C.~R.} \emph{et~al.}
\newblock \bibinfo{title}{{Interstellar Optics}}.
\newblock \emph{\bibinfo{journal}{\apj}} \textbf{\bibinfo{volume}{505}}, \bibinfo{pages}{928--940} (\bibinfo{year}{1998}).

\bibitem{2021ApJ...915...65M}
\bibinfo{author}{{Main}, R.} \emph{et~al.}
\newblock \bibinfo{title}{{Resolving the Emission Regions of the Crab Pulsar's Giant Pulses}}.
\newblock \emph{\bibinfo{journal}{\apj}} \textbf{\bibinfo{volume}{915}}, \bibinfo{pages}{65} (\bibinfo{year}{2021}).

\bibitem{2023ApJ...945..115L}
\bibinfo{author}{{Lin}, R.} \emph{et~al.}
\newblock \bibinfo{title}{{Resolving the Emission Regions of the Crab Pulsar's Giant Pulses. II. Evidence for Relativistic Motion}}.
\newblock \emph{\bibinfo{journal}{\apj}} \textbf{\bibinfo{volume}{945}}, \bibinfo{pages}{115} (\bibinfo{year}{2023}).
\newblock \eprint{2211.05209}.

\bibitem{2015Natur.528..523M}
\bibinfo{author}{{Masui}, K.} \emph{et~al.}
\newblock \bibinfo{title}{{Dense magnetized plasma associated with a fast radio burst}}.
\newblock \emph{\bibinfo{journal}{\nat}} \textbf{\bibinfo{volume}{528}}, \bibinfo{pages}{523--525} (\bibinfo{year}{2015}).
\newblock \eprint{1512.00529}.

\bibitem{2019MNRAS.483..971V}
\bibinfo{author}{{Vedantham}, H.~K.} \& \bibinfo{author}{{Phinney}, E.~S.}
\newblock \bibinfo{title}{{Radio wave scattering by circumgalactic cool gas clumps}}.
\newblock \emph{\bibinfo{journal}{\mnras}} \textbf{\bibinfo{volume}{483}}, \bibinfo{pages}{971--984} (\bibinfo{year}{2019}).
\newblock \eprint{1811.10876}.

\bibitem{2023arXiv230907256J}
\bibinfo{author}{{Jow}, D.~L.}, \bibinfo{author}{{Wu}, X.} \& \bibinfo{author}{{Pen}, U.-L.}
\newblock \bibinfo{title}{{Refractive lensing of scintillating FRBs by sub-parsec cloudlets in the multi-phase CGM}}.
\newblock \emph{\bibinfo{journal}{arXiv e-prints}} \bibinfo{pages}{arXiv:2309.07256} (\bibinfo{year}{2023}).
\newblock \eprint{2309.07256}.

\bibitem{2018ApJ...863...48C}
\bibinfo{author}{{CHIME/FRB Collaboration}} \emph{et~al.}
\newblock \bibinfo{title}{{The CHIME Fast Radio Burst Project: System Overview}}.
\newblock \emph{\bibinfo{journal}{\apj}} \textbf{\bibinfo{volume}{863}}, \bibinfo{pages}{48} (\bibinfo{year}{2018}).
\newblock \eprint{1803.11235}.

\bibitem{2021ApJ...910..147M}
\bibinfo{author}{{Michilli}, D.} \emph{et~al.}
\newblock \bibinfo{title}{{An Analysis Pipeline for CHIME/FRB Full-array Baseband Data}}.
\newblock \emph{\bibinfo{journal}{\apj}} \textbf{\bibinfo{volume}{910}}, \bibinfo{pages}{147} (\bibinfo{year}{2021}).
\newblock \eprint{2010.06748}.

\bibitem{2019ascl.soft10004S}
\bibinfo{author}{{Seymour}, A.}, \bibinfo{author}{{Michilli}, D.} \& \bibinfo{author}{{Pleunis}, Z.}
\newblock \bibinfo{title}{{DM\_phase: Algorithm for correcting dispersion of radio signals}}.
\newblock \bibinfo{howpublished}{Astrophysics Source Code Library, record ascl:1910.004} (\bibinfo{year}{2019}).

\bibitem{2002astro.ph..7156C}
\bibinfo{author}{{Cordes}, J.~M.} \& \bibinfo{author}{{Lazio}, T.~J.~W.}
\newblock \bibinfo{title}{{NE2001.I. A New Model for the Galactic Distribution of Free Electrons and its Fluctuations}}.
\newblock \emph{\bibinfo{journal}{arXiv e-prints}} \bibinfo{pages}{astro--ph/0207156} (\bibinfo{year}{2002}).
\newblock \eprint{astro-ph/0207156}.

\bibitem{2024RNAAS...8...17O}
\bibinfo{author}{{Ocker}, S.~K.} \& \bibinfo{author}{{Cordes}, J.~M.}
\newblock \bibinfo{title}{{NE2001p: A Native Python Implementation of the NE2001 Galactic Electron Density Model}}.
\newblock \emph{\bibinfo{journal}{Research Notes of the American Astronomical Society}} \textbf{\bibinfo{volume}{8}}, \bibinfo{pages}{17} (\bibinfo{year}{2024}).
\newblock \eprint{2401.05475}.

\bibitem{1990ARA&A..28..561R}
\bibinfo{author}{{Rickett}, B.~J.}
\newblock \bibinfo{title}{{Radio propagation through the turbulent interstellar plasma.}}
\newblock \emph{\bibinfo{journal}{\araa}} \textbf{\bibinfo{volume}{28}}, \bibinfo{pages}{561--605} (\bibinfo{year}{1990}).

\bibitem{2022ApJS..261...29C}
\bibinfo{author}{{CHIME Collaboration}} \emph{et~al.}
\newblock \bibinfo{title}{{An Overview of CHIME, the Canadian Hydrogen Intensity Mapping Experiment}}.
\newblock \emph{\bibinfo{journal}{\apjs}} \textbf{\bibinfo{volume}{261}}, \bibinfo{pages}{29} (\bibinfo{year}{2022}).
\newblock \eprint{2201.07869}.

\bibitem{2008PASA...25..184G}
\bibinfo{author}{{Gaensler}, B.~M.}, \bibinfo{author}{{Madsen}, G.~J.}, \bibinfo{author}{{Chatterjee}, S.} \& \bibinfo{author}{{Mao}, S.~A.}
\newblock \bibinfo{title}{{The Vertical Structure of Warm Ionised Gas in the Milky Way}}.
\newblock \emph{\bibinfo{journal}{\pasa}} \textbf{\bibinfo{volume}{25}}, \bibinfo{pages}{184--200} (\bibinfo{year}{2008}).
\newblock \eprint{0808.2550}.

\bibitem{2019ApJ...880..139V}
\bibinfo{author}{{Voit}, G.~M.}
\newblock \bibinfo{title}{{Ambient Column Densities of Highly Ionized Oxygen in Precipitation-limited Circumgalactic Media}}.
\newblock \emph{\bibinfo{journal}{\apj}} \textbf{\bibinfo{volume}{880}}, \bibinfo{pages}{139} (\bibinfo{year}{2019}).
\newblock \eprint{1811.04976}.

\bibitem{2003A&A...412...45P}
\bibinfo{author}{{Paturel}, G.} \emph{et~al.}
\newblock \bibinfo{title}{{HYPERLEDA. I. Identification and designation of galaxies}}.
\newblock \emph{\bibinfo{journal}{\aap}} \textbf{\bibinfo{volume}{412}}, \bibinfo{pages}{45--55} (\bibinfo{year}{2003}).

\bibitem{2018ApJ...866...54T}
\bibinfo{author}{{Tan}, C.~M.} \emph{et~al.}
\newblock \bibinfo{title}{{LOFAR Discovery of a 23.5 s Radio Pulsar}}.
\newblock \emph{\bibinfo{journal}{\apj}} \textbf{\bibinfo{volume}{866}}, \bibinfo{pages}{54} (\bibinfo{year}{2018}).
\newblock \eprint{1809.00965}.

\bibitem{2019MNRAS.485.4091M}
\bibinfo{author}{{Metzger}, B.~D.}, \bibinfo{author}{{Margalit}, B.} \& \bibinfo{author}{{Sironi}, L.}
\newblock \bibinfo{title}{{Fast radio bursts as synchrotron maser emission from decelerating relativistic blast waves}}.
\newblock \emph{\bibinfo{journal}{\mnras}} \textbf{\bibinfo{volume}{485}}, \bibinfo{pages}{4091--4106} (\bibinfo{year}{2019}).
\newblock \eprint{1902.01866}.

\bibitem{2019ApJ...876L...6P}
\bibinfo{author}{{Philippov}, A.}, \bibinfo{author}{{Uzdensky}, D.~A.}, \bibinfo{author}{{Spitkovsky}, A.} \& \bibinfo{author}{{Cerutti}, B.}
\newblock \bibinfo{title}{{Pulsar Radio Emission Mechanism: Radio Nanoshots as a Low-frequency Afterglow of Relativistic Magnetic Reconnection}}.
\newblock \emph{\bibinfo{journal}{\apjl}} \textbf{\bibinfo{volume}{876}}, \bibinfo{pages}{L6} (\bibinfo{year}{2019}).
\newblock \eprint{1902.07730}.

\bibitem{2012ApJ...758....7G}
\bibinfo{author}{{Gwinn}, C.~R.} \emph{et~al.}
\newblock \bibinfo{title}{{Size of the Vela Pulsar's Emission Region at 18 cm Wavelength}}.
\newblock \emph{\bibinfo{journal}{\apj}} \textbf{\bibinfo{volume}{758}}, \bibinfo{pages}{7} (\bibinfo{year}{2012}).
\newblock \eprint{1208.0040}.

\bibitem{2008ARA&A..46..127H}
\bibinfo{author}{{Hester}, J.~J.}
\newblock \bibinfo{title}{{The Crab Nebula : an astrophysical chimera.}}
\newblock \emph{\bibinfo{journal}{\araa}} \textbf{\bibinfo{volume}{46}}, \bibinfo{pages}{127--155} (\bibinfo{year}{2008}).

\bibitem{2017Natur.541...58C}
\bibinfo{author}{{Chatterjee}, S.} \emph{et~al.}
\newblock \bibinfo{title}{{A direct localization of a fast radio burst and its host}}.
\newblock \emph{\bibinfo{journal}{\nat}} \textbf{\bibinfo{volume}{541}}, \bibinfo{pages}{58--61} (\bibinfo{year}{2017}).
\newblock \eprint{1701.01098}.

\bibitem{2017ApJ...834L...8M}
\bibinfo{author}{{Marcote}, B.} \emph{et~al.}
\newblock \bibinfo{title}{{The Repeating Fast Radio Burst FRB 121102 as Seen on Milliarcsecond Angular Scales}}.
\newblock \emph{\bibinfo{journal}{\apjl}} \textbf{\bibinfo{volume}{834}}, \bibinfo{pages}{L8} (\bibinfo{year}{2017}).
\newblock \eprint{1701.01099}.

\bibitem{2022Natur.606..873N}
\bibinfo{author}{{Niu}, C.~H.} \emph{et~al.}
\newblock \bibinfo{title}{{A repeating fast radio burst associated with a persistent radio source}}.
\newblock \emph{\bibinfo{journal}{\nat}} \textbf{\bibinfo{volume}{606}}, \bibinfo{pages}{873--877} (\bibinfo{year}{2022}).
\newblock \eprint{2110.07418}.

\bibitem{2023ApJ...958L..19B}
\bibinfo{author}{{Bhandari}, S.} \emph{et~al.}
\newblock \bibinfo{title}{{Constraints on the Persistent Radio Source Associated with FRB 20190520B Using the European VLBI Network}}.
\newblock \emph{\bibinfo{journal}{\apjl}} \textbf{\bibinfo{volume}{958}}, \bibinfo{pages}{L19} (\bibinfo{year}{2023}).
\newblock \eprint{2308.12801}.

\bibitem{2018ApJ...868L...4M}
\bibinfo{author}{{Margalit}, B.} \& \bibinfo{author}{{Metzger}, B.~D.}
\newblock \bibinfo{title}{{A Concordance Picture of FRB 121102 as a Flaring Magnetar Embedded in a Magnetized Ion-Electron Wind Nebula}}.
\newblock \emph{\bibinfo{journal}{\apjl}} \textbf{\bibinfo{volume}{868}}, \bibinfo{pages}{L4} (\bibinfo{year}{2018}).
\newblock \eprint{1808.09969}.

\bibitem{2020ApJ...895....7Y}
\bibinfo{author}{{Yang}, Y.-P.}, \bibinfo{author}{{Li}, Q.-C.} \& \bibinfo{author}{{Zhang}, B.}
\newblock \bibinfo{title}{{Are Persistent Emission Luminosity and Rotation Measure of Fast Radio Bursts Related?}}
\newblock \emph{\bibinfo{journal}{\apj}} \textbf{\bibinfo{volume}{895}}, \bibinfo{pages}{7} (\bibinfo{year}{2020}).
\newblock \eprint{2001.10761}.

\bibitem{2024MNRAS.527.7568O}
\bibinfo{author}{{Ocker}, S.~K.} \emph{et~al.}
\newblock \bibinfo{title}{{Pulsar scintillation through thick and thin: bow shocks, bubbles, and the broader interstellar medium}}.
\newblock \emph{\bibinfo{journal}{\mnras}} \textbf{\bibinfo{volume}{527}}, \bibinfo{pages}{7568--7587} (\bibinfo{year}{2024}).
\newblock \eprint{2309.13809}.

\bibitem{2022ApJ...925..109L}
\bibinfo{author}{{Levkov}, D.~G.}, \bibinfo{author}{{Panin}, A.~G.} \& \bibinfo{author}{{Tkachev}, I.~I.}
\newblock \bibinfo{title}{{Propagation Effects in the FRB 20121102A Spectra}}.
\newblock \emph{\bibinfo{journal}{\apj}} \textbf{\bibinfo{volume}{925}}, \bibinfo{pages}{109} (\bibinfo{year}{2022}).
\newblock \eprint{2010.15145}.

\bibitem{1991tstm.book.....B}
\bibinfo{author}{{Brockwell}, P.~J.} \& \bibinfo{author}{{Davis}, R.~A.}
\newblock \emph{\bibinfo{title}{{Time series: Theory and methods}}} (\bibinfo{year}{1991}).

\bibitem{2020ApJ...904..104R}
\bibinfo{author}{{Reardon}, D.~J.} \emph{et~al.}
\newblock \bibinfo{title}{{Precision Orbital Dynamics from Interstellar Scintillation Arcs for PSR J0437-4715}}.
\newblock \emph{\bibinfo{journal}{\apj}} \textbf{\bibinfo{volume}{904}}, \bibinfo{pages}{104} (\bibinfo{year}{2020}).
\newblock \eprint{2009.12757}.

\bibitem{2017MNRAS.470.2659G}
\bibinfo{author}{{Geyer}, M.} \emph{et~al.}
\newblock \bibinfo{title}{{Scattering analysis of LOFAR pulsar observations}}.
\newblock \emph{\bibinfo{journal}{\mnras}} \textbf{\bibinfo{volume}{470}}, \bibinfo{pages}{2659--2679} (\bibinfo{year}{2017}).
\newblock \eprint{1706.04205}.

\bibitem{2022ApJ...931...87O}
\bibinfo{author}{{Ocker}, S.~K.} \emph{et~al.}
\newblock \bibinfo{title}{{The Large Dispersion and Scattering of FRB 20190520B Are Dominated by the Host Galaxy}}.
\newblock \emph{\bibinfo{journal}{\apj}} \textbf{\bibinfo{volume}{931}}, \bibinfo{pages}{87} (\bibinfo{year}{2022}).
\newblock \eprint{2202.13458}.

\bibitem{2023MNRAS.525.5653S}
\bibinfo{author}{{Sammons}, M.~W.} \emph{et~al.}
\newblock \bibinfo{title}{{Two-screen scattering in CRAFT FRBs}}.
\newblock \emph{\bibinfo{journal}{\mnras}} \textbf{\bibinfo{volume}{525}}, \bibinfo{pages}{5653--5668} (\bibinfo{year}{2023}).
\newblock \eprint{2305.11477}.

\bibitem{2013ApJ...776..125M}
\bibinfo{author}{{Macquart}, J.-P.} \& \bibinfo{author}{{Koay}, J.~Y.}
\newblock \bibinfo{title}{{Temporal Smearing of Transient Radio Sources by the Intergalactic Medium}}.
\newblock \emph{\bibinfo{journal}{\apj}} \textbf{\bibinfo{volume}{776}}, \bibinfo{pages}{125} (\bibinfo{year}{2013}).
\newblock \eprint{1308.4459}.

\bibitem{2020MNRAS.498..651B}
\bibinfo{author}{{Beniamini}, P.} \& \bibinfo{author}{{Kumar}, P.}
\newblock \bibinfo{title}{{What does FRB light-curve variability tell us about the emission mechanism?}}
\newblock \emph{\bibinfo{journal}{\mnras}} \textbf{\bibinfo{volume}{498}}, \bibinfo{pages}{651--664} (\bibinfo{year}{2020}).
\newblock \eprint{2007.07265}.

\bibitem{2004ApJ...605..759B}
\bibinfo{author}{{Bhat}, N.~D.~R.}, \bibinfo{author}{{Cordes}, J.~M.}, \bibinfo{author}{{Camilo}, F.}, \bibinfo{author}{{Nice}, D.~J.} \& \bibinfo{author}{{Lorimer}, D.~R.}
\newblock \bibinfo{title}{{Multifrequency Observations of Radio Pulse Broadening and Constraints on Interstellar Electron Density Microstructure}}.
\newblock \emph{\bibinfo{journal}{\apj}} \textbf{\bibinfo{volume}{605}}, \bibinfo{pages}{759--783} (\bibinfo{year}{2004}).
\newblock \eprint{astro-ph/0401067}.

\bibitem{2022MNRAS.510.4654B}
\bibinfo{author}{{Beniamini}, P.}, \bibinfo{author}{{Kumar}, P.} \& \bibinfo{author}{{Narayan}, R.}
\newblock \bibinfo{title}{{Faraday depolarization and induced circular polarization by multipath propagation with application to FRBs}}.
\newblock \emph{\bibinfo{journal}{\mnras}} \textbf{\bibinfo{volume}{510}}, \bibinfo{pages}{4654--4668} (\bibinfo{year}{2022}).
\newblock \eprint{2110.00028}.

\bibitem{2008MNRAS.388.1214W}
\bibinfo{author}{{Walker}, M.~A.}, \bibinfo{author}{{Koopmans}, L.~V.~E.}, \bibinfo{author}{{Stinebring}, D.~R.} \& \bibinfo{author}{{van Straten}, W.}
\newblock \bibinfo{title}{{Interstellar holography}}.
\newblock \emph{\bibinfo{journal}{\mnras}} \textbf{\bibinfo{volume}{388}}, \bibinfo{pages}{1214--1222} (\bibinfo{year}{2008}).
\newblock \eprint{0801.4183}.

\bibitem{2010ApJ...708..232B}
\bibinfo{author}{{Brisken}, W.~F.} \emph{et~al.}
\newblock \bibinfo{title}{{100 {\ensuremath{\mu}}as Resolution VLBI Imaging of Anisotropic Interstellar Scattering Toward Pulsar B0834+06}}.
\newblock \emph{\bibinfo{journal}{\apj}} \textbf{\bibinfo{volume}{708}}, \bibinfo{pages}{232--243} (\bibinfo{year}{2010}).
\newblock \eprint{0910.5654}.

\bibitem{2024ApJ...962...57S}
\bibinfo{author}{{Serafin Nadeau}, T.} \emph{et~al.}
\newblock \bibinfo{title}{{A Cacophony of Echoes from Daily Monitoring of the Crab Pulsar at Jodrell Bank}}.
\newblock \emph{\bibinfo{journal}{\apj}} \textbf{\bibinfo{volume}{962}}, \bibinfo{pages}{57} (\bibinfo{year}{2024}).
\newblock \eprint{2310.07007}.

\bibitem{1998Obs...118..201G}
\bibinfo{author}{{Goodwin}, S.~P.}, \bibinfo{author}{{Gribbin}, J.} \& \bibinfo{author}{{Hendry}, M.~A.}
\newblock \bibinfo{title}{{The relative size of the Milky Way}}.
\newblock \emph{\bibinfo{journal}{The Observatory}} \textbf{\bibinfo{volume}{118}}, \bibinfo{pages}{201--208} (\bibinfo{year}{1998}).

\bibitem{2020yCat.1350....0G}
\bibinfo{author}{{Gaia Collaboration}}.
\newblock \bibinfo{title}{{VizieR Online Data Catalog: Gaia EDR3 (Gaia Collaboration, 2020)}}.
\newblock \bibinfo{howpublished}{VizieR On-line Data Catalog: I/350. Originally published in: 2021A\&A...649A...1G; doi:10.5270/esa-1ug} (\bibinfo{year}{2020}).

\bibitem{2017ApJ...843...15W}
\bibinfo{author}{{Walker}, M.~A.} \emph{et~al.}
\newblock \bibinfo{title}{{Extreme Radio-wave Scattering Associated with Hot Stars}}.
\newblock \emph{\bibinfo{journal}{\apj}} \textbf{\bibinfo{volume}{843}}, \bibinfo{pages}{15} (\bibinfo{year}{2017}).
\newblock \eprint{1705.00964}.

\bibitem{2015ExA....39..259K}
\bibinfo{author}{{Keimpema}, A.} \emph{et~al.}
\newblock \bibinfo{title}{{The SFXC software correlator for very long baseline interferometry: algorithms and implementation}}.
\newblock \emph{\bibinfo{journal}{Experimental Astronomy}} \textbf{\bibinfo{volume}{39}}, \bibinfo{pages}{259--279} (\bibinfo{year}{2015}).
\newblock \eprint{1502.00467}.

\bibitem{whitney_2010_ivs}
\bibinfo{author}{{Whitney}, A.}, \bibinfo{author}{{Kettenis}, M.}, \bibinfo{author}{{Phillips}, C.} \& \bibinfo{author}{{Sekido}, M.}
\newblock \bibinfo{title}{{VLBI Data Interchange Format (VDIF)}}.
\newblock In \bibinfo{editor}{{Navarro}, R.} \emph{et~al.} (eds.) \emph{\bibinfo{booktitle}{Sixth International VLBI Service for Geodesy and Astronomy. Proceedings from the 2010 General Meeting}}, \bibinfo{pages}{192--196} (\bibinfo{year}{2010}).

\bibitem{2003ASSL..285..109G}
\bibinfo{author}{{Greisen}, E.~W.}
\newblock \bibinfo{title}{{AIPS, the VLA, and the VLBA}}.
\newblock In \bibinfo{editor}{{Heck}, A.} (ed.) \emph{\bibinfo{booktitle}{Information Handling in Astronomy - Historical Vistas}}, vol. \bibinfo{volume}{285} of \emph{\bibinfo{series}{Astrophysics and Space Science Library}}, \bibinfo{pages}{109} (\bibinfo{year}{2003}).

\bibitem{1994BAAS...26..987S}
\bibinfo{author}{{Shepherd}, M.~C.}, \bibinfo{author}{{Pearson}, T.~J.} \& \bibinfo{author}{{Taylor}, G.~B.}
\newblock \bibinfo{title}{{DIFMAP: an interactive program for synthesis imaging.}}
\newblock In \emph{\bibinfo{booktitle}{Bulletin of the American Astronomical Society}}, vol.~\bibinfo{volume}{26}, \bibinfo{pages}{987--989} (\bibinfo{year}{1994}).

\bibitem{2020PASP..132c5001L}
\bibinfo{author}{{Lacy}, M.} \emph{et~al.}
\newblock \bibinfo{title}{{The Karl G. Jansky Very Large Array Sky Survey (VLASS). Science Case and Survey Design}}.
\newblock \emph{\bibinfo{journal}{\pasp}} \textbf{\bibinfo{volume}{132}}, \bibinfo{pages}{035001} (\bibinfo{year}{2020}).
\newblock \eprint{1907.01981}.

\bibitem{2011PASA...28....1V}
\bibinfo{author}{{van Straten}, W.} \& \bibinfo{author}{{Bailes}, M.}
\newblock \bibinfo{title}{{DSPSR: Digital Signal Processing Software for Pulsar Astronomy}}.
\newblock \emph{\bibinfo{journal}{\pasa}} \textbf{\bibinfo{volume}{28}}, \bibinfo{pages}{1--14} (\bibinfo{year}{2011}).
\newblock \eprint{1008.3973}.

\bibitem{2020MNRAS.497.1661A}
\bibinfo{author}{{Agarwal}, D.}, \bibinfo{author}{{Aggarwal}, K.}, \bibinfo{author}{{Burke-Spolaor}, S.}, \bibinfo{author}{{Lorimer}, D.~R.} \& \bibinfo{author}{{Garver-Daniels}, N.}
\newblock \bibinfo{title}{{FETCH: A deep-learning based classifier for fast transient classification}}.
\newblock \emph{\bibinfo{journal}{\mnras}} \textbf{\bibinfo{volume}{497}}, \bibinfo{pages}{1661--1674} (\bibinfo{year}{2020}).
\newblock \eprint{1902.06343}.

\bibitem{2003ApJ...596.1142C}
\bibinfo{author}{{Cordes}, J.~M.} \& \bibinfo{author}{{McLaughlin}, M.~A.}
\newblock \bibinfo{title}{{Searches for Fast Radio Transients}}.
\newblock \emph{\bibinfo{journal}{\apj}} \textbf{\bibinfo{volume}{596}}, \bibinfo{pages}{1142--1154} (\bibinfo{year}{2003}).
\newblock \eprint{astro-ph/0304364}.

\bibitem{2022ApJ...927...59L}
\bibinfo{author}{{Lanman}, A.~E.} \emph{et~al.}
\newblock \bibinfo{title}{{A Sudden Period of High Activity from Repeating Fast Radio Burst 20201124A}}.
\newblock \emph{\bibinfo{journal}{\apj}} \textbf{\bibinfo{volume}{927}}, \bibinfo{pages}{59} (\bibinfo{year}{2022}).
\newblock \eprint{2109.09254}.

\bibitem{2021NatAs...5..594N}
\bibinfo{author}{{Nimmo}, K.} \emph{et~al.}
\newblock \bibinfo{title}{{Highly polarized microstructure from the repeating FRB 20180916B}}.
\newblock \emph{\bibinfo{journal}{Nature Astronomy}} \textbf{\bibinfo{volume}{5}}, \bibinfo{pages}{594--603} (\bibinfo{year}{2021}).
\newblock \eprint{2010.05800}.

\end{thebibliography}


\end{methods}
\begin{addendum}

    
    
    \item
    We thank Benito Marcote for help with the EVN observations, Ramesh Karuppusamy for help with the pulsar backend recording at Effelsberg, Dylan Jow for helpful discussions about anisotropic screens, Jim Cordes and Stella Ocker for answering questions about NE2001 and Jason Hessels for helpful discussions.

    K.N. is an MIT Kavli Fellow. Z.P. was a Dunlap Fellow and is supported by an NWO Veni fellowship (VI.Veni.222.295). P.B. is supported by a grant (no. 2020747) from the United States-Israel Binational Science Foundation (BSF), Jerusalem, Israel by a grant (no. 1649/23) from the Israel Science Foundation and by a grant (no. 80NSSC 24K0770)  from the NASA astrophysics theory program. PK is supported in part by an NSF grant AST-2009619 and a NASA grant 80NSSC24K0770. M.W.S. acknowledges support from the Trottier Space Institute Fellowship program. A.P.C. is a Vanier Canada Graduate Scholar. The Dunlap Institute is funded through an endowment established by the David Dunlap family and the University of Toronto. B.M.G. acknowledges the support of the Natural Sciences and Engineering Research Council of Canada (NSERC) through grant RGPIN-2022-03163, and of the Canada Research Chairs program. V.M.K. holds the Lorne Trottier Chair in Astrophysics \& Cosmology, a Distinguished James McGill Professorship, and receives support from an NSERC Discovery grant (RGPIN 228738-13), from an R. Howard Webster Foundation Fellowship from CIFAR, and from the FRQNT CRAQ. C.~L. is supported by NASA through the NASA Hubble Fellowship grant HST-HF2-51536.001-A awarded by the Space Telescope Science Institute, which is operated by the Association of Universities for Research in Astronomy, Inc., under NASA contract NAS5-26555. K.W.M. holds the Adam J. Burgasser Chair in Astrophysics and is supported by NSF grants (2008031, 2018490).  A.P. is funded by the NSERC Canada Graduate Scholarships -- Doctoral program. A.B.P. is a Banting Fellow, a McGill Space Institute~(MSI) Fellow, and a Fonds de Recherche du Quebec -- Nature et Technologies~(FRQNT) postdoctoral fellow. K.S. is supported by the NSF Graduate Research Fellowship Program.FRB research at UBC is supported by an NSERC Discovery Grant and by the Canadian Institute for Advanced Research.  The baseband recording system on CHIME/FRB is funded in part by a CFI John R. Evans Leaders Fund grant to IHS.

    We would like to thank the directors and staff at the various participating EVN stations for allowing us to use their facilities and running the observations. The European VLBI Network is a joint facility of independent European, African, Asian, and North American radio astronomy institutes. Scientific results from data presented in this publication are derived from the following EVN project code: RN002.

    
    \item[Competing Interests] The authors declare that they have no competing financial interests.
    
    \item[Correspondence] 
\end{addendum}

\clearpage
\newpage

\section*{Extended Data}

\setcounter{figure}{0}
\setcounter{table}{0}

\captionsetup[table]{name={\bf Extended Data Table}}

\begin{figureED}

\centerline{\resizebox{100mm}{!}{\includegraphics{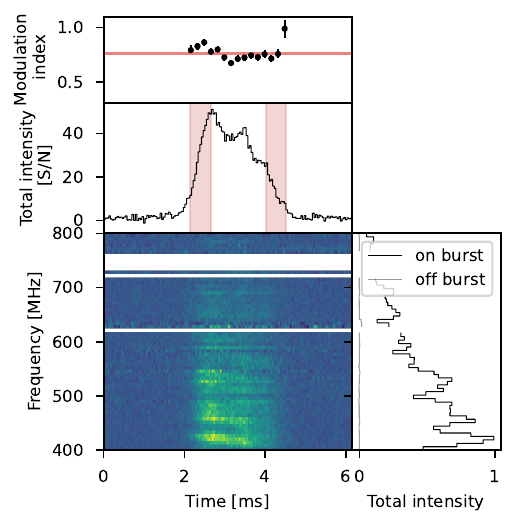}}}
\caption{FRB~20221022A burst dynamic spectrum (bottom left panel), profile (middle panel), spectrum (right panel) and modulation index (top panel). The burst is dedispersed to a dispersion measure\cite{2024arXiv240209304M} of $116.8371\,\mathrm{pc}\,\mathrm{cm}^{-3}$ and is plotted with time and frequency resolution $40.96\,\upmu$s and $6.2$\,MHz, respectively. The rise and decay time are highlighted using the shaded red regions in the middle panel. Both the on-burst time-averaged spectrum and off-burst spectrum are shown in the bottom right panel. For each 163.84\,$\upmu$s time bin, we compute the ACF (Equation\,\ref{eq:acf}) across frequency (ACF is computed for spectra with a frequency resolution of 24\,kHz), and measure the modulation index as the height of the Lorentzian fit to the ACF around zero lag. We only plot modulation indices for 163.84\,$\upmu$s time bins that have a S/N$>8$. The mean of the measured time resolved modulation indices for the $124$\,kHz scintillation scale is shown with the red line, and is measured to be $\bar{m}=0.76\pm0.06$, consistent with the frequency-resolved modulation index measured for this scintillation scale. } \label{fig:ds}
\end{figureED}
\clearpage

\begin{figureED}
   \centerline{\resizebox{180mm}{!}{\includegraphics{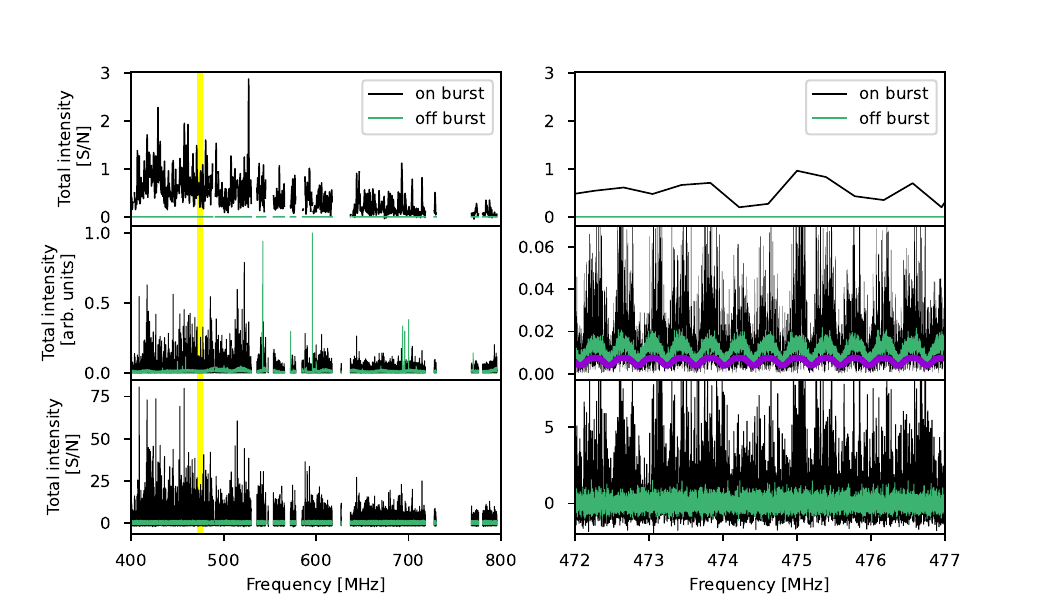}}}
\caption{On-burst and off-burst spectra across the CHIME observing band from 400--800\,MHz (left panels). A zoom-in around 472--477\,MHz (the yellow bar in the left panels) is plotted in the right panels. Top panels are the spectra of the baseband data with frequency resolution 0.39\,MHz (1024 channels across the entire observing band). The upchannelised spectra (frequency resolution: 0.76\,kHz) are shown in the middle panels \emph{before} correcting for the scalloping introduced by the FFT. The model we use to correct the scalloping is shown in purple in the zoom-in panel. The bottom panels show the spectra after correcting for the upchannelisation scalloping, and applying additional RFI masking.  \label{fig:spec}}
\end{figureED}
\clearpage

\begin{figureED}
\centerline{\resizebox{150mm}{!}{\includegraphics{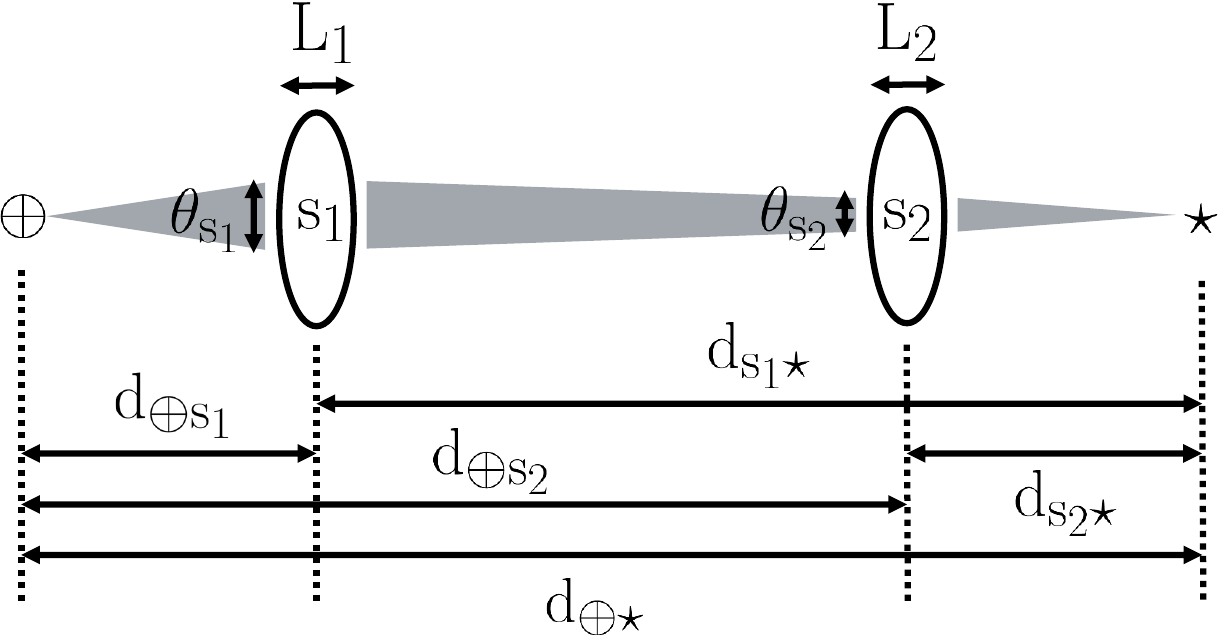}}}
\caption{Diagram of a two-screen lensing setup with a source ($\star$), screen nearest the source ($\mathrm{s}_2$), screen nearest the observer ($\mathrm{s}_1$) and observer ($\oplus$), all the distances, $\mathrm{d}$, between any two elements and angular broadening due to scattering, $\theta$.} 
\label{fig:twoscreen}
\end{figureED}

\clearpage
\begin{figureED}
\centerline{\resizebox{100mm}{!}{\includegraphics{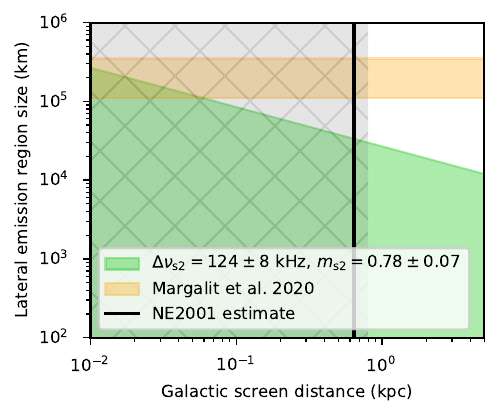}}}
\caption{The lateral emission region size as it depends on the Galactic screen distance, $d_{\oplus\mathrm{s}1}$, through the relationship shown on Figure\,\ref{fig:emission_size} and the two-screen constraint in Equation\,\ref{eq:2scr}. The green shaded region shows the allowable lateral emission region sizes and Galactic screen distance combinations for our measured scintillation parameters at $600$\,MHz: $\Delta\nu_{\mathrm{s}2}=124$\,kHz and $m_{\mathrm{s}2}=0.78$. The black vertical line indicates the assumed $d_{\oplus\mathrm{s}1}=0.64$\,kpc from NE2001. The orange shaded region shows the emission region sizes estimated for the synchrotron maser shock model in Ref.\cite{2020MNRAS.494.4627M}. The grey hatched region shows the parameter space we ruled out based on the apparent diameter of the host galaxy and the two-screen constraint in Equation\,\ref{eq:2scr} (see Figure\,\ref{fig:emission_size}).} 
\label{fig:galscreen}
\end{figureED}
\clearpage

\begin{figureED}
\centerline{\resizebox{180mm}{!}{\includegraphics{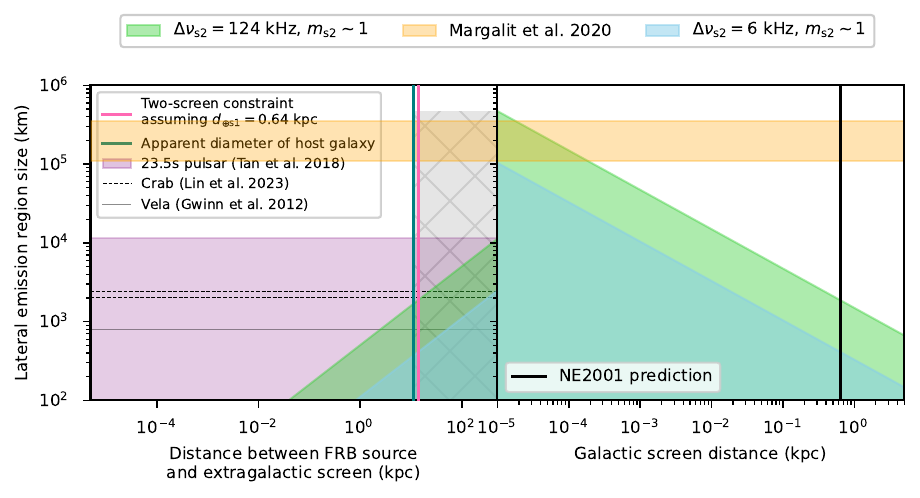}}}
\caption{Lateral emission region size constraints for the other cases we consider: case (b) in the text refers to the extragalactic screen having a scintillation bandwidth of $124$\,kHz at $600$\,MHz and modulation index $\sim1$ (green shaded region); and case (c) for the extragalactic screen having a decorrelation bandwidth of $6$\,kHz at $600$\,MHz and modulation index $\sim1$ (blue shaded region). The left panel is the same as Figure\,\ref{fig:emission_size} (case (a)) for different scintillation measurements (case (b) and case (c)), and the right panel is the same as Extended Data Figure\,\ref{fig:galscreen} for the additional cases considered. } 
\label{fig:othercases}
\end{figureED}

\end{document}